\documentclass[11pt,a4paper]{article}
\usepackage{jcappub}
\usepackage{slashed}
\usepackage{braket}
\usepackage{hyperref}
\usepackage{float}
\usepackage[normalem]{ulem}

\newcommand{\aeq}{a_\mathrm{eq}}
\newcommand{\ain}{a_\mathrm{in}}

\newcommand{\cambcode}{{\tt camb}}
\newcommand{\classcode}{{\tt class}}
\newcommand{\DerB}{\Delta^{\mathrm{(B)}}}
\newcommand{\DerC}{\Delta^{\mathrm{(C)}}}
\newcommand{\dKron}{\delta^\mathrm{(K)}}
\newcommand{\Ec}{{\mathcal E}}
\newcommand{\ellmax}{\ell_\mathrm{max}}
\newcommand{\gadgetcode}{{\tt Gadget-2}}
\newcommand{\Hc}{{\mathcal H}}
\newcommand{\Hco}{{\mathcal H}_0}
\newcommand{\Ombo}{\Omega_{\mathrm{b}0}}
\newcommand{\Omcbo}{\Omega_{\mathrm{cb}0}}
\newcommand{\Omgo}{\Omega_{\gamma 0}}
\newcommand{\Omm}{\Omega_{\mathrm{m}}}
\newcommand{\Ommo}{\Omega_{\mathrm{m}0}}
\newcommand{\Omno}{\Omega_{\nu 0}}
\newcommand{\Omnmasslesso}{\Omega_{\nu\mathrm{(massless),}0}}

\newcommand{\Pleg}{{\mathcal P}}

\title{The cosmic neutrino background as a collection of fluids in large-scale structure simulations}

\author[a]{Joe Zhiyu Chen,}
\author[a]{Amol Upadhye,} 
\author[a]{Yvonne Y.~Y.~Wong}

\affiliation[a]{Sydney Consortium for Particle Physics and Cosmology, School of Physics, The University of New South Wales, 
Sydney NSW 2052, Australia} 

\emailAdd{zhiyu.chen@unsw.edu.au}
\emailAdd{a.upadhye@unsw.edu.au}
\emailAdd{yvonne.y.wong@unsw.edu.au}

\abstract{
  A significant challenge for modelling the massive neutrino as a hot dark matter is its large velocity dispersion.  In this work, we investigate and implement a multi-fluid perturbation theory that treats the cosmic neutrino population as a collection of fluids with a broad range of bulk velocities. These fluids respond linearly to the clustering of cold matter, which may be linear and described by standard linear perturbation theory, or non-linear, described using either higher-order perturbation theory or $N$-body simulations.  We verify that such an alternative treatment of neutrino perturbations agrees closely with state-of-the-art neutrino linear response calculations in terms of power spectrum and bispectrum predictions.  Combining multi-fluid neutrino linear response with a non-linear calculation for the cold matter clustering, we find for a reference $\nu \Lambda$CDM cosmology with neutrino mass sum $\sum m_\nu = 0.93$~eV an enhancement of the small-scale neutrino power by an order of magnitude relative to a purely linear calculation.  
  The corresponding clustering enhancement in the cold matter, however, is a modest $\sim 0.05\%$.   Importantly, our multi-fluid approach uniquely enables  us to identify that the slowest-moving 25\% of the neutrino population clusters strongly enough to warrant a non-linear  treatment.  Such a precise calculation of neutrino clustering  on small scales accompanied by fine-grained velocity information would be invaluable for experiments such as PTOLEMY that probe the local neutrino density and velocity in the solar neighbourhood.
  }

\begin{document}
	
\begin{flushright}
	{\large \tt CPPC-2020-08}
\end{flushright}		
	
\maketitle

\section{Introduction}
\label{sec:introduction}

Cosmology is within reach of a measurement of the sum of neutrino masses, $\sum m_\nu$, one of the final unmeasured parameters of the Standard Model of particle physics.  The upper bound on $\sum m_\nu$ in the simplest cosmological model --- currently at $\sum m_\nu\lesssim 0.12$~eV (95\%)~\cite{Aghanim:2018eyx} --- is rapidly converging on the lower bound from laboratory oscillation experiments, which can measure mass differences but not the sum~\cite{deSalas:2017kay,Esteban:2018azc}.  Nevertheless, powerful motivations exist for continuing to consider larger neutrino masses.  Simple extensions to the minimal cosmological model, such as a time-variation in the dark energy density accelerating the expansion of the universe, weaken the cosmological upper bound on $\sum m_\nu$ by a factor of about three~\cite{Upadhye:2017hdl}.  Neutrino constraints based upon the scale-dependent suppression of galaxy clustering suffer from systematic uncertainties due to scale-dependent galaxy bias and difficult-to-model baryonic effects.  Moreover, persistent tensions and anomalies in cosmology and particle physics, such as the $H_0$ tension between early and late probes~\cite{Alam:2016hwk,Riess:2019cxk,Wong:2019kwg}, the $\sigma_8$~tension~\cite{Abbott:2017wau}, and the LSND/MiniBooNE/reactor anomalies~\cite{Abazajian:2012ys}
 may be explained by a sterile neutrino with mass $\sim 1$~eV that is not part of the Standard Model.

Additionally, in recent years an array of techniques has emerged for probing neutrino phenomena other than their small-scale suppression of cold dark matter (CDM) and baryon clustering.  Neutrino clustering around CDM halos leads to a characteristic scale-dependence of the halo bias at the free-streaming scale below which neutrino clustering is suppressed~\cite{LoVerde:2013lta,LoVerde:2014pxa,Chiang:2017vuk,Chiang:2018laa}.  Neutrinos have a velocity relative to CDM and baryons, meaning that a stream of neutrinos flowing past collapsed CDM+baryon structures will form a ``wake'' on the other side~\cite{Zhu:2013tma,Zhu:2014qma}.  Halos in neutrino-rich environments will capture neutrinos more efficiently, leading to an observable change in their mass function~\cite{Yu:2016yfe}.  Neutrinos also lead to long-range correlations in the spins of galaxies~\cite{Yu:2018llx}.
Thus, understanding neutrino clustering itself, and not just its impact on the clustering of CDM and baryons, is essential for deriving robust neutrino mass constraints from cosmological observations.

However, the large velocity dispersion expected of a massive relic neutrino population presents a daunting challenge for the numerical investigation of neutrinos' gravitational clustering. 
Neutrinos at late times are neither radiation, whose clustering may be understood purely perturbatively, nor cold matter, which may be realised as particles with a well-defined velocity field in an $N$-body simulation.  A particle neutrino simulation must instead follow the full six-dimensional phase space of neutrinos~\cite{Brandbyge:2009ce,Viel:2010bn,Castorina:2015bma,Banerjee:2016zaa,Banerjee:2018bxy}, a significant challenge at a time when accurate simulation of the three-dimensional CDM+baryon phase space requires the most powerful supercomputers on the planet.    Some existing approximations to get around this problem in simulations include representing the neutrino population as a fluid on a grid~\cite{Dakin:2017idt,Tram:2018znz}, and linear perturbation theory/linear response~\cite{Brandbyge:2008js,Agarwal:2010mt,AliHaimoud:2012vj,Upadhye:2013ndm,Mummery:2017lcn,Liu:2017now,Inman:2020oda}.

The same large neutrino velocity dispersion also plagues {\it perturbative} fluid treatments of the  neutrino population, because the continuity and Euler equations of fluid dynamics apply to a fluid with a well-defined density and velocity at each spatial point~\cite{Ma:1993xs,Shoji:2010hm,Fuhrer:2014zka,Elbers:2020lbn}.  In this regard, a significant advance came from Dupuy and Bernardeau, who, instead of treating the entire neutrino population as a single fluid, subdivided the population into a set of discrete flows, each with a different but well-defined bulk velocity field, thereby allowing the fluid equations to be applied to each individual flow~\cite{Dupuy:2013jaa,Dupuy:2014vea,Dupuy:2015ega}.  In addition to rendering the system amenable to standard fluid perturbation techniques, their approach reduces the dimensionality of the problem from six to five, as the equations of motion are symmetric under a rotation of the Fourier-space wave vector about the neutrino flow direction.  Dupuy and Bernardeau implemented their method in linear perturbation theory and showed it to agree with the widely-used Boltzmann hierarchy~\cite{Ma:1995ey} employed in standard linear cosmological Boltzmann codes such as~\cambcode{}~\cite{Lewis:1999bs} and \classcode{}~\cite{Lesgourgues:2011re,Lesgourgues:2011rh,Blas:2011rf}.  A similar perturbation theory for constant-speed neutrino shells was developed and implemented in an $N$-body simulation in reference~\cite{Inman:2020oda}.

Since we are primarily interested in the formation of cosmic structure at late times, we develop a non-relativistic multi-fluid neutrino perturbation theory based upon the relativistic theory of Dupuy and Bernardeau.  Dependences of the density and velocity perturbations on the angle between the Fourier vector and particle momentum are expanded in Legendre polynomials and truncated at finite order.  After verifying the convergence and accuracy of this linear perturbation theory, we combine it with a non-linear perturbative calculation for the CDM+baryon fluid, thereby allowing each of the many neutrino fluids to respond linearly to non-linear CDM+baryon clustering.  Next, we incorporate multi-fluid neutrinos into a \gadgetcode{} $N$-body simulation for a more realistic calculation of neutrino clustering.  

Equipped with this multi-fluid linear response calculation for neutrinos, we systematically study the enhancement of neutrino and CDM+baryon clustering due to neutrino response.  In a reference model with $\sum m_\nu = 0.93$~eV, neutrino clustering is enhanced by over an order of magnitude relative to linear theory predictions
at the smallest scale accessible to our simulation, $k \approx 2~h/$Mpc.  Meanwhile, CDM+baryon clustering is only modestly enhanced by $\lesssim 0.1\%$ in the power spectrum and bispectrum.  Finally, we use the fine-grained velocity information provided by the multi-fluid perturbation theory to quantify the fraction of the neutrino population that clusters with a dimensionless power spectrum $\Delta^2$ exceeding $\approx 0.1$, a regime in which our linear perturbative treatment is expected to break down.  We find that over 25\% of the neutrinos enter this non-linear regime in our reference model.  When $\sum m_\nu$ is reduced to $0.5$~eV, a value compatible with data constraints if the dark energy equation of state is allowed to vary~\cite{Upadhye:2017hdl}, a smaller but still significant  $5\%$ of the neutrinos enter the non-linear regime.  This result motivates a future hybrid $N$-body simulation in which the slowest perturbative neutrino fluids are converted into particles as they reach the non-linear regime, in the manner of~\cite{Brandbyge:2009ce,Bird:2018all}.

The outline for this article is as follows.  In section~\ref{sec:relativistic_multi-fluid_PT} we briefly review the relativistic multi-fluid perturbation theory of Dupuy and Bernardeau.  We proceed in section~\ref{sec:multi-fluid_PT_in_subhorizon_NR} to develop a non-relativistic version of such a multi-fluid perturbation theory, which we subject to a battery of tests in section~\ref{sec:testing_neutrino_linear_response}.  In section~\ref{sec:linear_response_to_non-linear_clustering} we combine it with a non-linear perturbation theory for the CDM+baryon fluid, and estimate the impact of neutrino linear response on clustering.  We incorporate this multi-fluid treatment into the \gadgetcode{} N-body simulation code in section~\ref{sec:mflr_in_N-body_simulations}, and use it to provide more accurate calculations of the impact of neutrino linear response, before concluding in section~\ref{sec:conclusions}.

\section{Relativistic multi-fluid perturbation theory}
\label{sec:relativistic_multi-fluid_PT}

The cosmological neutrino population is not a perfect fluid because it has a velocity dispersion. However,  references~\cite{Dupuy:2013jaa,Dupuy:2014vea,Dupuy:2015ega} discovered a method to decompose this population into subsets with uniform velocities at zeroth order in perturbation theory.  In the limit of a large number of subsets, the collection of such subsets approaches the velocity distribution of the cosmological neutrino population.  The advantage of this approach is that each of these subsets of particles is a fluid described at all points by a density and a momentum, and thus amenable to standard fluid perturbative treatments.

We provide in this section a brief review of this multi-fluid perturbation theory~\cite{Dupuy:2013jaa,Dupuy:2014vea,Dupuy:2015ega}.  
Working within the conformal Newtonian gauge, the line element is given by
\begin{equation}
  {\rm d}s^2
  =
  a^2(\eta)
  \left[
    -(1+2\Phi){\rm d}\eta^2 + (1-2\Psi)\left|{\rm d} \vec{x}\right|^2
  \right],
\end{equation}
and the conformal Hubble expansion rate is $\Hc(\eta) := {\rm d}\log(a)/{\rm d}\eta$.  It is convenient to change our time variable to $s := \ln(a/\ain)$ for a constant $\ain$, so that the conformal Hubble rate can be equivalently expressed as
\begin{eqnarray}
\frac{\Hc^2(s)}{\Hco^2}
&=&
\sum \Omega_{0} \, \Ec(s),
\label{e:Hubble}
\\
\frac{\Hc'(s)}{\Hco}
&=&
\frac{1}{2}
\frac{\Hc^2(s)}{\Hco^2}
\sum \Omega_{0} \, \Ec'(s),
\label{e:dHubble}
\end{eqnarray}
where primes ($'$) denote derivatives with respect to~$s$, and~$\Hco$ is its present-day value of~$\Hc$.  In writing equations~\eqref{e:Hubble} and \eqref{e:dHubble} we have assumed the universe to consist of various forms of matter and energy each with a present-day energy density fraction~$\Omega_{0}$. Each energy density evolves with time as $\Omega(s) = \Omega_{0}\,  \Ec(s) \Hco^2/\Hc^2$, where the function $\Ec(s)$ encapsulates the fluid's thermodynamic and kinematic properties.  A perfect fluid at rest with equation of state $w = P/\rho$ has $\Ec =a^{-(1+3w)}$, while an isotropic collection of fluids each with uniform speed $v$ has $\Ec = a^{-1} (1-v^2)^{-1/2}$.
Since particles of mass $m$ with comoving momenta $\vec \tau$ have speeds $v = (1 + m^2 a^2 / \tau^2)^{-1/2}$, the latter expression for $\Ec$ smoothly interpolates between the $w=0$ and $1/3$ limits of the former, perfect fluid expression as $\tau$ is increased from zero to infinity.


\subsection{Neutrinos as a collection of fluids} 
\label{subsec:fluidcollection}

Consider a universe containing only a relic population of neutrinos of mass $m_\nu$, subdivided into an infinite number of fluids according to some property $\vec{\tau}$.
At position~$\vec{x}$ and time~$s$, each of these neutrino fluids is characterised by  a number density $n(s,\vec x,\vec{\tau})$ and a 4-momentum $P^\mu(s,\vec x,\vec{\tau})$, whose  spatial means at a given time~$s$ are $\bar n(s,\vec{\tau})$ and $\bar{P}^\mu(s,\vec{\tau})$ respectively.  The mean number density of one such fluid evolves with the scale factor as $\bar n(s,\vec{\tau}) \propto a^{-3}(s)$ as usual.  On the other hand, the spatial part of the mean 4-momentum with {\it upper} Lorentz indices scales as $\bar P^i(s,\vec{\tau}) \propto a^{-2}(s)$, while its {\it lower} index version $\bar P_i(s,\vec{\tau})$ is constant in time.  The latter leads us to identify $\tau_i := \bar P_i(s,\vec{\tau})$, and it follows that the mean coordinate 3-velocity of the fluid is $\vec v(s,\vec{\tau}) = - \vec \tau / \xi(s,\tau)$, where $\tau := |\vec\tau|$ and $\xi(s,\tau) := -\sqrt{m_\nu^2 a^2(s) + \tau^2}$.

At early times when the spatial inhomogeneities are small, we expect $P_i(s,\vec x,\vec{\tau}) \rightarrow \tau_i$.  Then, subdividing the neutrino population into fluids by $\vec{\tau}$ is equivalent to labelling these fluids by their initial comoving 3-momenta.  Each fluid thus characterised is unique, and its subsequent evolution under the influence of spatial inhomogeneities can be treated with standard fluid perturbation theory.
Reference~\cite{Dupuy:2013jaa} has derived such a  linear perturbation theory for a neutrino fluid characterised by an initial momentum~$\vec \tau$, which   forms the starting point for our analysis.  In Fourier space,  the equations of motion for a  fluid's number density contrast $\delta (s, \vec{k},\vec{\tau}) := n(s,\vec{k},\vec{\tau})/\bar{n}(s,\vec{\tau}) - 1$ and the momentum divergence $\theta_P(s, \vec{k},\vec{\tau}) := i k^i P_i (s, \vec{k},\vec{\tau})$ are respectively~\cite{Dupuy:2013jaa}
\begin{eqnarray}
  \delta'
  &=&
  i\frac{\vec k \cdot \vec \tau}{\Hc \xi} \delta
  + \left(1 - \frac{(\vec k \cdot \vec \tau)^2}{k^2 \xi^2}\right)\theta_P
  + 3\Psi'  
  - i\frac{\vec k \cdot \vec \tau}{\Hc \xi}
  \left[\left(1+\frac{\tau^2}{\xi^2}\right)\Psi - \Phi\right],
  \label{e:delta_DB13}
  \\
  \theta_P'
  &=&
  i\frac{\vec k \cdot \vec \tau}{\Hc \xi} \theta_P
  - \frac{\xi k^2}{\Hc} \Phi
  - \frac{\tau^2 k^2}{\xi \Hc} \Psi.
  \label{e:thetaP_DB13}
\end{eqnarray}
The key difference between these equations and standard linearised fluid equations for, e.g., CDM, is that the initial fluid momentum~$\vec \tau$ now also enters the picture.  Thus, this multi-fluid treatment of neutrinos is inherently Lagrangian in momentum space.

Two observations about equations~\eqref{e:delta_DB13} and \eqref{e:thetaP_DB13} simplify their analysis.  Firstly, the direction of $\vec \tau$ appears only through its inner product with the Fourier vector, $\mu := \vec k \cdot \vec \tau / (k \tau)$.  This together with the fact that the initial conditions are also independent of the absolute direction of $\vec \tau$ means that
 we need not consider the full set of allowed $\vec \tau$ when representing the neutrino population as a set of discrete fluids.   It suffices to deal only with $\tau$ and $\mu$, and modelling relic neutrinos can proceed first by partitioning the population by $\tau$  into $N_\tau$  fluids, followed by approximating each these $N_\tau$ fluids as having $N_\mu$ angular degrees of freedom, in the form of either a discrete set of $\mu$ as in reference~\cite{Dupuy:2013jaa}, or a finite-dimensional subspace of the set of all functions of $\mu$.
 
 Secondly, different fluids interact only through the gravitational potentials $\Phi(s,k)$ and $\Psi(s,k)$.  For an infinite number of fluids, each with density contrast~$\delta(s,k,\tau,\mu)$  and velocity perturbations~$\theta_P(s,k,\tau,\mu)$, the potentials  in the conformal Newtonian gauge are~\cite{Dupuy:2014vea}
 \begin{eqnarray}
   k^2 \Psi + 3 \Hc^2 (\Psi' + \Phi)
   &=&
   -4 \pi G a^2 \delta\rho,
   \\
   k^2(\Psi' + \Phi)
   &=&
   -4 \pi G a^2 \delta U,
   \\
   \delta\rho
   :=
   -\frac{1}{2} \int_{-1}^1 {\rm d}\mu \, \delta T^0_0
   &=&
   \bar \rho_\mathrm{crit}(\tau)
   \int_0^\infty {\rm d} \tau \, \Omega_\nu(s,\tau)
   \int_{-1}^{1} \frac{{\rm d} \mu}{2}
   \left( \delta - \frac{\tau \mu}{k \xi^2}\theta_P
   + \frac{\tau^2}{\xi^2}\Psi\right),
   \\
   \delta U
   :=
   -\frac{1}{2} \int_{-1}^1 {\rm d}\mu \frac{i k_j}{\Hc} \delta T^0_j
   &=&
   \bar \rho_\mathrm{crit}(\tau)
   \int_0^\infty {\rm d} \tau \, \Omega_\nu(s,\tau)
   \int_{-1}^{1} \frac{{\rm d} \mu}{2}
   \left( \frac{\theta_P + k\tau\mu \delta }{\Hc \xi}  \right),
 \end{eqnarray}
 where $\Omega_\nu(s,\tau) = \Omega_{\nu 0}(\tau) \, \Ec(s,\tau) \Hc_0^2/\Hc^2$, with $\Ec(s,\tau) = |\xi| / (m_\nu a^2)$.  At a given~$\tau$, these potentials depend on the quantities $\delta_{\ell=0}(s,k,\tau) = \frac{1}{2} \int_{-1}^1 {\rm d}\mu \, \delta$,  $\delta_{\ell=1}(s,k,\tau) = \frac{1}{2} \int_{-1}^1 {\rm d}\mu \, \mu \delta$, $\theta_{P, \ell=0}(s,k,\tau) =  \frac{1}{2} \int_{-1}^1 {\rm d}\mu \, \theta_P$, and $\theta_{P, \ell=1}(s,k,\tau) =  \frac{1}{2} \int_{-1}^1 {\rm d}\mu \, \mu \theta_P$, which are the Legendre monopoles and dipoles of the fluid perturbations.  This motivates us to represent the $N_\mu$ angular degrees of freedom using  a Legendre polynomial expansion.

We describe first in the following how we sample $N_\tau$ fluids following the method of reference~\cite{Dupuy:2013jaa}. Our Legendre representation of the fluids' angular degrees of freedom differs from the approach of~\cite{Dupuy:2013jaa}, and we defer its discussion to section~\ref{subsec:subhorizon_non-relativistic_fluid_equations}.


\subsection{Sampling neutrino momenta}
\label{subsec:neutrino_momenta}

 We sample the initial neutrino fluid momenta~$\tau$  from an ultrarelativistic Fermi--Dirac distribution.   Defining  $x := \tau / (a T_\nu)$, where  
 $T_\nu(s) = T_{\nu,0} / a(s)$, with $T_{\nu,0} = 1.95$~K, is the neutrino temperature, 
 the (time-indepemdent) mean comoving number density of the full neutrino population  is given up to a constant multiplicative factor by
\begin{equation}
\begin{aligned}
  \bar n_\mathrm{comoving}
 & \propto \, T^3_{\nu,0}
  \int_0^\infty {\rm d}x \, \frac{x^2}{1 + \exp(x)} \\
  & =\,  \frac{3 \zeta(3)}{2} T^3_{\nu,0} ,
  \end{aligned}
\end{equation}
where $\zeta(3)$ is a Riemann zeta function.
The cumulative probability  ${\mathfrak P}(\tau)$ of finding a neutrino with momentum less than $\tau$ is therefore
\begin{equation}
  {\mathfrak P}(\tau)
  =
  \frac{2}{3\zeta(3)}
  \int_0^{\tau / T_{\nu,0}} {\rm d}x \, \frac{x^2}{1+\exp(x)}.
\end{equation}
This is by definition a monotonically increasing function. It may therefore be inverted, thereby defining $\tau(\mathfrak P)$ for $0 \leq \mathfrak P \leq 1$.

We choose to divide the neutrino population into $N_\tau$ bins of equal number densities, implying that all neutrino density fractions $\Omega_\alpha(s)$ are equal in the non-relativistic limit.  Then, for each neutrino fluid $\alpha=0, \ldots, N_\tau-1$ in our collection of $N_\tau$ fluids, a convenient choice of $\tau_\alpha$ is $\tau_\alpha = \tau({\mathfrak P}_\alpha)$ for ${\mathfrak P}_\alpha = (\alpha + \frac{1}{2}) / N_\tau$.  With this choice, the parameter $N_\tau$ determines the momentum resolution within the neutrino population. Thus, although the multi-fluid method allows for an arbitrary binning, which may be tailored for specific applications, we do not investigate this possibility further. 

\section{Multi-fluid perturbation theory in the subhorizon non-relativistic limit}
\label{sec:multi-fluid_PT_in_subhorizon_NR}

 References~\cite{Dupuy:2013jaa,Dupuy:2014vea} describe a general treatment of multi-fluid neutrinos capable of replacing the Boltzmann hierarchy in codes such as \cambcode{} and \classcode{}~\cite{Lewis:1999bs,Lewis:2002nc,Seljak:1996is,Zaldarriaga:1997va,Blas:2011rf,Lesgourgues:2011re,Lesgourgues:2011rh}.  Our interest here, on the other hand, is subhorizon structure formation at late times when neutrinos have become non-relativistic, and 
we are particularly interested in combining this multi-fluid description of neutrino perturbations with a non-linear model of CDM+baryon clustering in the so-called linear response approach.  The goal of this section, therefore, is to adapt the multi-fluid perturbation theory to the non-relativistic, subhorizon limit, and to test this theory numerically against other, more well-established  perturbative methods.

In the following, we shall always assume that the neutrino population has already been discretised in $\tau$ as per section~\ref{subsec:neutrino_momenta}, so that 
 the continuous variables of section~\ref{subsec:fluidcollection} need now be replaced with their discrete counterparts: $\tau \to \tau_{\alpha}$,
$\theta_P(\tau) \to \theta_{P,\alpha}$, $\delta(\tau) \to \delta_{\alpha}$, $v(\tau) \to v_\alpha$, $\xi(\tau) \to \xi_\alpha$, $\Omega(\tau) \to \Omega_\alpha$, etc.,  where the subscript $\alpha$ labels a discrete neutrino fluid.  Where confusion is unlikely to arise, we shall  omit writing out the time variable~$s$.


\subsection{Subhorizon non-relativistic fluid equations} 
\label{subsec:subhorizon_non-relativistic_fluid_equations}

We begin by replacing  in equation~(\ref{e:thetaP_DB13}) the momentum divergence by the velocity divergence for each neutrino fluid $\alpha$, $\theta_\alpha := \theta_{P,\alpha} / (\Hc \xi_{\alpha})$.  The resulting equation of motion is
\begin{equation}
  \theta_\alpha'
  =
  -\left(1 - v_\alpha^2 + \frac{\Hc'}{\Hc}\right)\theta_\alpha
  - \frac{ik\mu v_\alpha}{\Hc} \theta_\alpha
  - \frac{k^2}{\Hc^2}(\Phi + v_\alpha^2\Psi).
  \label{e:eom_theta_full}
\end{equation}
In the non-relativistic limit we may discard terms of order $v_\alpha^2$, so that
\begin{equation}
    \theta_\alpha'
  =
  -\left(1 + \frac{\Hc'}{\Hc}\right)\theta_\alpha
  - \frac{ik\mu v_\alpha}{\Hc} \theta_\alpha
  - \frac{k^2}{\Hc^2}\Phi.
  \label{e:theta_nr_sub}
\end{equation}
Similarly, recast in terms of $\theta_{\alpha}$ equation~(\ref{e:delta_DB13}) becomes
\begin{equation}
  \delta_\alpha'
  =
  -\frac{ik\mu v_\alpha}{\Hc} \delta_\alpha
  + (1-\mu^2 v_\alpha^2)\theta_\alpha
  + 3\Psi'
  + \frac{ik\mu v_\alpha}{\Hc} \left[(1+v_\alpha^2)\Psi - \Phi\right],
  \label{e:eom_delta_full}
\end{equation}
which reduces further in the subhorizon non-relativistic limit to
\begin{equation}
  \delta_\alpha'
  =
  -\frac{ik\mu v_\alpha}{\Hc} \delta_\alpha
  + \theta_\alpha
  \label{e:delta_nr_sub}
\end{equation}
if we neglect all terms of order $v_\alpha^2$ as well as  gravitational potentials not multiplied by $k^2/\Hc^2$.

Departing from reference~\cite{Dupuy:2013jaa}, we now expand $\delta_\alpha(k,\mu)$ and $\theta_\alpha(k,\mu)$ in Legendre polynomials $\Pleg_\ell(\mu)$.  For an arbitrary function~$X(\mu)$, we define the Legendre coefficients $X_\ell$ via
\begin{equation}
\begin{aligned}
  X(\mu) &= \, \sum_{\ell=0}^\infty (-i)^\ell \Pleg_\ell(\mu) X_\ell, \\
  X_\ell &=\,  \frac{i^\ell }{2} (2\ell+1) \int_{-1}^1 d\mu \Pleg_\ell(\mu) X(\mu).
  \label{eq:legendre}
  \end{aligned}
\end{equation}
This particular  choice of definition follows from the structure of our equations of motion~\eqref{e:theta_nr_sub} and \eqref{e:delta_nr_sub}, which couple terms odd and even in $\mu$ only through factors proportional to $i\mu$; the factors of $i$ in the definition~\eqref{eq:legendre} therefore ensure that the  Legendre coefficients $\delta_{\alpha,\ell}(k)$ and $\theta_{\alpha,\ell}(k)$ initialised to real values will remain real under time evolution. 
The corresponding Legendre-expanded equations of motion are
\begin{eqnarray}
  \delta_{\alpha,\ell}'
  &=&
  \theta_{\alpha,\ell}
  + \frac{k v_\alpha}{\Hc}\left(
    \frac{\ell}{2\ell-1} \delta_{\alpha,\ell-1}
    -\frac{\ell+1}{2\ell+3} \delta_{\alpha,\ell+1} \right),
    \label{e:dnu_eom_ell}
  \\
  \theta_{\alpha,\ell}'
  &=&
  -\dKron_{\ell 0} \frac{k^2}{\Hc^2} \Phi
  -\left(1+\frac{\Hc'}{\Hc}\right)\theta_{\alpha,\ell}
  +\frac{k v_\alpha}{\Hc}\left(
    \frac{\ell}{2\ell-1} \theta_{\alpha,\ell-1}
    -\frac{\ell+1}{2\ell+3} \theta_{\alpha,\ell+1} \right),
    \label{e:tnu_eom_ell}
\end{eqnarray}
where $\dKron_{ij}$ is the Kronecker delta function.

This system of equations couples each $\ell \geq 0$ to $\ell+1$, resulting in an infinite system of equations that needs to be truncated in order to limit the decompositions of  $\delta_\alpha(k,\mu)$ and $\theta_\alpha(k,\mu)$  to a finite number $N_\mu$ of Legendre polynomials. 
Reference~\cite{Ma:1995ey} warns against truncating this system at a finite $\ellmax=N_\mu - 1$ by merely setting all higher coefficients to zero;  doing so would amount to imposing a reflecting boundary condition at $\ellmax$, thereby sending power that ought to flow towards infinite $\ell$ back towards $\ell =0$.  Thus, we must approximate $\delta_{\alpha,\ell}$ and $\theta_{\alpha,\ell}$ at $\ell = \ellmax + 1$.

In order to do this, note that the terms multiplying $k v_\alpha/\Hc$ in equations~\eqref{e:dnu_eom_ell} and~\eqref{e:tnu_eom_ell} in the limit of large $\ell$ approach finite-difference approximations to the first derivatives of $\delta_{\alpha,\ell}$ and $\theta_{\alpha,\ell}$ with respect to $\ell$.  We make this explicit by defining $f_\ell(x) = x \delta_{\alpha,x\ell} / (x\ell + 1/2)$.  Then, the centered second-order finite-difference approximation to the first derivative is $\DerC_{2,\epsilon} f_\ell(x) = [f_\ell(x+\epsilon)-f_\ell(x-\epsilon)]/(2\epsilon)$, with the actual derivative ${\rm d}f_\ell(x)/{\rm d}x = \DerC_{2,\epsilon} f_\ell(x) + {\mathcal O}(\epsilon^2)$.  For $x=1$ and $\epsilon=1/\ell$, we find
\begin{equation}
  \DerC_{2,\frac{1}{\ell}} f_\ell(1)
  =
  \frac{\ell+1}{2\ell+3} \delta_{\alpha,\ell+1}
  - \frac{\ell-1}{2\ell-1} \delta_{\alpha,\ell-1},
\end{equation}
which can be easily rearranged to give
\begin{equation}
\frac{\ell}{2\ell-1} \delta_{\alpha,\ell-1} 
-\frac{\ell+1}{2\ell+3} \delta_{\alpha,\ell+1}
=
\frac{1}{2\ell-1} \delta_{\alpha,\ell-1}- \DerC_{2,\frac{1}{\ell}} f_\ell(1),
\end{equation}
i.e., precisely the term in parenthesis in equation~(\ref{e:dnu_eom_ell}) now written in terms of $\DerC_{2,1/\ell} f_\ell(1)$.

Next, we replace the centered derivative $\DerC_{2,\epsilon} f_\ell(x)$ at $\ell=N_\mu-1$ by the backwards derivative $\DerB_{2,\epsilon} f_\ell(x) = [f_\ell(x-2\epsilon]) - 4f_\ell(x-\epsilon) + 3f_\ell(x)]/(2\epsilon)$, which also approximates  ${\rm d}f_\ell(x)/{\rm d}x$ to ${\mathcal O}(\epsilon^2) \approx {\mathcal O}(N_\mu^{-2})$. The key is that $\DerB_{2,1/\ellmax} f_\ell(1)$ depends only upon $\delta_{\alpha,\ell}$ for $\ell \leq \ellmax$.  This replacement amounts to approximating
\begin{equation}
  \frac{\ellmax+1}{2\ellmax+3} \delta_{\alpha,\ellmax+1}
  \approx
  \frac{3\ellmax}{2\ellmax+1} \delta_{\alpha,\ellmax}
  - \frac{3(\ellmax-1)}{2\ellmax-1} \delta_{\alpha,\ellmax-1}
  + \frac{(\ellmax-2)}{2\ellmax-3} \delta_{\alpha,\ellmax-2}.
  \label{e:d_truncation}
\end{equation}
Similar arguments apply to the truncation of equation~(\ref{e:tnu_eom_ell}).  In practice, we find $N_\mu \approx 10$ to be sufficient for percent-level accuracy in $\delta_{\alpha,\ellmax}$ and $\theta_{\alpha,\ellmax}$, which themselves should be subdominant at late times.  This conclusion is consistent with that of reference~\cite{Dupuy:2013jaa}, which found that high-accuracy neutrino calculations do not require a high resolution of the $\mu$-dependence of the perturbations.

Finally, in the subhorizon limit $\Hc \ll k$, the Newtonian-gauge potentials are equal, i.e., $\Psi = \Phi$, and   Poisson's equation relates them to the density perturbations of the various constituents of the universe:
\begin{equation}
 k^2 \Phi(s,k)
  =
  -\frac{3}{2} \Hc^2(s) \left(\Omega_{\rm cb}(s)  \delta_{\rm cb}(s,k)+   
    \sum_{\alpha=0} ^{N_\tau-1}
  \Omega_\alpha(s) \delta_{\alpha,\ell=0} (s,k) \right).
  \label{e:Poisson}
\end{equation}
Here, $\Omega_{\rm cb}(s)$ and $\delta_{\rm cb}(s,k)$ denote respectively the time-dependent energy density fraction and density contrast of the 
 combined CDM+baryon fluid,  $\delta_{\alpha,\ell=0}$ is the monopole of the $\alpha$th neutrino fluid density contrast, and the summation is over all $N_\tau$ neutrino fluids weighted by~$\Omega_\alpha(s)$, which we have chosen to be equal for non-relativistic neutrinos.

Then, to  summarise, our multi-fluid neutrino perturbation theory uses the subhorizon non-relativistic equations of motion~\eqref{e:dnu_eom_ell} and~\eqref{e:tnu_eom_ell} truncated as per equation~(\ref{e:d_truncation}) at $\ellmax = N_\mu-1$, and with a gravitational potential given by Poisson's equation~(\ref{e:Poisson}).


\subsubsection{The $v_\alpha =0$ limit: CDM+baryon equations of motion}

Before moving on, let us consider briefly the $v_\alpha=0$ limit of equations~\eqref{e:dnu_eom_ell} and \eqref{e:tnu_eom_ell}.  Formally setting $v_\alpha$  to zero immediately leads to the decoupling of the different $\ell$ moments in both equations.  Since the presence of the Kronecker delta $\delta_{\ell 0}^{(K)}$ ensures that only the monopole ($\ell=0$) equations contain a gravitational source term, it suffices to consider only these equations, i.e.,
\begin{eqnarray}
\delta_{\ell=0}'
&=&
\theta_{\ell=0}\, ,
\label{e:cdm+baryon-delta}
\\
\theta_{\ell=0}'
&=&
- \frac{k^2}{\Hc^2} \Phi
-\left(1+\frac{\Hc'}{\Hc}\right)\theta_{\ell=0}\, ,
\label{e:cdm+baryon-theta}
\end{eqnarray}
where we have also dropped the $\alpha$ fluid labels.

Equations~\eqref{e:cdm+baryon-delta} and~\eqref{e:cdm+baryon-theta} are formally identically the standard linearised fluid equations for cold matter (see, e.g.,~\cite{Bernardeau:2001qr}).  That the monopole equations of motion for a neutrino fluid with $v_\alpha=0$ also describe the time evolution of CDM+baryons is of course expected, since by definition a cold fluid has no velocity dispersion.  In the following, whenever the linear CDM+baryon density contrast~$\delta_{\rm cb}$ and velocity divergence~$\theta_{\rm cb}$ are called for, they shall be computed from equations~\eqref{e:cdm+baryon-delta} and~\eqref{e:cdm+baryon-theta} with
 $\delta_{\ell=0} \to \delta_{\rm cb}$ and $\theta_{\ell=0} \to \theta_{\rm cb}$,  unless otherwise specified.


\subsection{Initial conditions}
\label{subsec:initial_conditions}

Closed-form growing-mode solutions to equations~\eqref{e:dnu_eom_ell}, \eqref{e:tnu_eom_ell}, \eqref{e:Poisson}, \eqref{e:cdm+baryon-delta} and \eqref{e:cdm+baryon-theta} in a generic $\nu \Lambda$CDM universe are not known.  
  However, since the growing mode is an attractor solution, an approximate solution suffices for the purpose of setting initial conditions.  We describe our approximation below.  In all cases, we initialise at the scale factor $\ain = 10^{-3}$.

Consider first a universe containing only cold matter and radiation. The growing-mode solution for linear CDM+bayron perturbations in such a universe is well known and given by
\begin{eqnarray}
  \delta_{\rm cb}(s) & =&  a(s) + \frac{2}{3}\aeq ,
  \label{e:cb_growing-mode-delta} \\
 \theta_{\rm cb}(s) & = & \delta'(s) = a, 
   \label{e:cb_growing-mode-theta} 
  \end{eqnarray}
where $\aeq$ is the scale factor at matter--radiation equality.  To adapt these expressions into approximate growing-mode initial conditions for $\delta_{\rm cb}$ and $\theta_{\rm cb}$ in a realistic universe that also contains massive neutrinos, we note first of all that at $a_{\rm in} = 10^{-3}$ massive neutrinos are typically neither radiation nor cold matter
but are transitioning the former to the latter.  Then,
to approximate this transition effect, 
we set 
\begin{equation}
\aeq = \frac{\Omgo + \Omnmasslesso}{\Omcbo},
\end{equation}
where $\Omnmasslesso$ is the $z=0$ density fraction corresponding to massless neutrinos with temperature $T_{\nu,0} = 1.95$~K.  Thus the radiation component counts only those constituents that are truly massless, while only CDM+baryons contribute to the cold matter.  In other words, the energy density in massive neutrinos does not feature in the evaluation of $\aeq$ for the purpose of setting initial conditions via equations~\eqref{e:cb_growing-mode-delta} and~\eqref{e:cb_growing-mode-theta}.

Meanwhile in the neutrino sector, we note that given a CDM+baryon density contrast~$\delta_\mathrm{cb}$,  the total neutrino monopole density contrast, $\delta_{\nu, \ell=0}: \sum_{\alpha=0}^{N_\tau-1} \Omega_\alpha \delta_{\alpha, \ell=0}/\sum_{\alpha=0}^{N_\tau-1} \Omega_\alpha$, can be approximated by~\cite{Ringwald:2004np,Wong:2008ws}
\begin{equation}
  \frac{\delta_{\nu, 0}(s,k)}{\delta_\mathrm{cb}(s,k)}
  =
  \frac{k_\mathrm{FS}^2 (1-f_\nu)}{(k+k_\mathrm{FS})^2 - f_\nu k_\mathrm{FS}^2},
  \label{e:dnu_tot_IC}
 \end{equation}
where  $f_\nu := \Omega_{\rm 0}/\Omega_{\nu 0}$ is the neutrino mass fraction, and 
\begin{equation}  
  k_\mathrm{FS}(s)
 =
  \frac{\Hc}{c_{\nu}} \sqrt{\frac{3}{2} \Omm(s)}
  =
  a \Hc \frac{m_\nu}{T_{\nu,0}} \sqrt{\frac{\ln(2)}{\zeta(3)}\Omm(s)}
  \label{e:kfs_tot}
\end{equation}
is the time-dependent neutrino free-streaming wave number, with $c_{\nu}$ denoting a characteristic thermal speed of the relic neutrino population.  Inspired by these approximations, we initialise the monopole ($\ell=0$) of the neutrino fluid $\alpha$ to 
\begin{eqnarray}
  \delta_{\alpha, 0}(s,k)
& = &
  \frac{k_{\mathrm{FS},\alpha}^2 (1-f_\nu)}{(k+k_{\mathrm{FS},\alpha})^2 - f_\nu k_{\mathrm{FS},\alpha}^2} \, \delta_\mathrm{cb}(s,k),
  \label{e:dnu_alpha_IC}
  \\
 \theta_{\alpha,0} (s,k) &=& \delta_{\alpha, 0}'(s,k),
\end{eqnarray}  
where the $\alpha$-dependent free-streaming wave number is now given by
 \begin{equation}   
  k_{\mathrm{FS},\alpha}(s)
  =
  \frac{\Hc}{v^{\rm NR}_\alpha} \sqrt{\frac{3}{2} \Omm(s)}
  =
  \frac{m_\nu a \Hc}{\tau_\alpha} \sqrt{\frac{3}{2}\Omm(s)}\, ,
  \label{e:kfs_alpha}
\end{equation}
with the characteristic thermal speed $c_\nu$ now replaced with the fluid's non-relativistic velocity $v_\alpha^{\rm NR} :=\tau_\alpha / (m_\nu a)$. 
All  $\ell>0$ terms of the neutrino density contrast and velocity divergence are initialised to zero at $\ain$.


\section{Testing  multi-fluid linear perturbations}
\label{sec:testing_neutrino_linear_response}

Having written down in section~\ref{sec:multi-fluid_PT_in_subhorizon_NR}
	a multi-fluid linear perturbation theory for massive neutrinos in the subhorizon, non-relativistic limit, we now test the approximations made in the multi-fluid approach and show that the theory agrees with standard calculations of the linear neutrino density contrast. 
We restrict our attention in this section to a fully linearised set-up,  wherein the linear CDM+baryon density contrast $\delta_{\rm cb}$ and velocity divergence~$\theta_{\rm cb}$ are computed from equations~\eqref{e:cdm+baryon-delta} and~\eqref{e:cdm+baryon-theta}.
 The case of non-linear CDM+baryon clustering will be discussed in sections~\ref{sec:linear_response_to_non-linear_clustering} and~\ref{sec:mflr_in_N-body_simulations}.

\begin{table}[t]
	\begin{center}
		\footnotesize 
		\begin{tabular}{lcc}
			\hline
			\hline
			Parameter & Symbol & Value \\ 
			\hline
			Total matter density &  $\Ommo h^2$ &  0.1335 \\ 
			Baryon energy density & $\Ombo h^2$ &   0.02258 \\
			Neutrino energy density & $\Omno h^2$ &   0.01 \\
			Reduced Hubble parameter &  $h$ & 0.71 \\
			Scalar spectral index & $n_s$ & 0.963\\
			RMS linear matter density fluctuation on $8\, h$/Mpc & $\sigma_8$ & 0.8 \\
			Optical depth to reionisation &  $\tau_{\rm re}$  & 0.09296 \\
			\hline
			\hline
		\end{tabular}
	\end{center}
	\caption{Cosmological parameter values of the reference $\nu \Lambda$CDM model used in this work.  The chosen neutrino energy density corresponds to $\sum m_\nu = 0.93$~eV, which we assume to be equally distributed amongst three families.\label{t:nuLCDM_reference}}
\end{table}

 For the remainder of the work, unless otherwise noted, we shall use the reference $\nu\Lambda$CDM cosmology specified in table~\ref{t:nuLCDM_reference}.
We further assume three equal-mass neutrino species, implying that each neutrino mass is $m_\nu \approx 0.31$~eV. Though such a mass is much larger than current minimal bounds derived from extending the $\Lambda$CDM model with only~$\sum m_\nu$, we note that analyses that also allow two dark energy parameters to vary yield constraints consistent with $m_\nu \lesssim 0.2$~eV at the $2\sigma$~level~\cite{Upadhye:2017hdl}.

\subsection{Convergence tests}

\subsubsection{Series truncation}
\begin{figure}[t]
  \begin{center}
    \includegraphics[width=150mm]{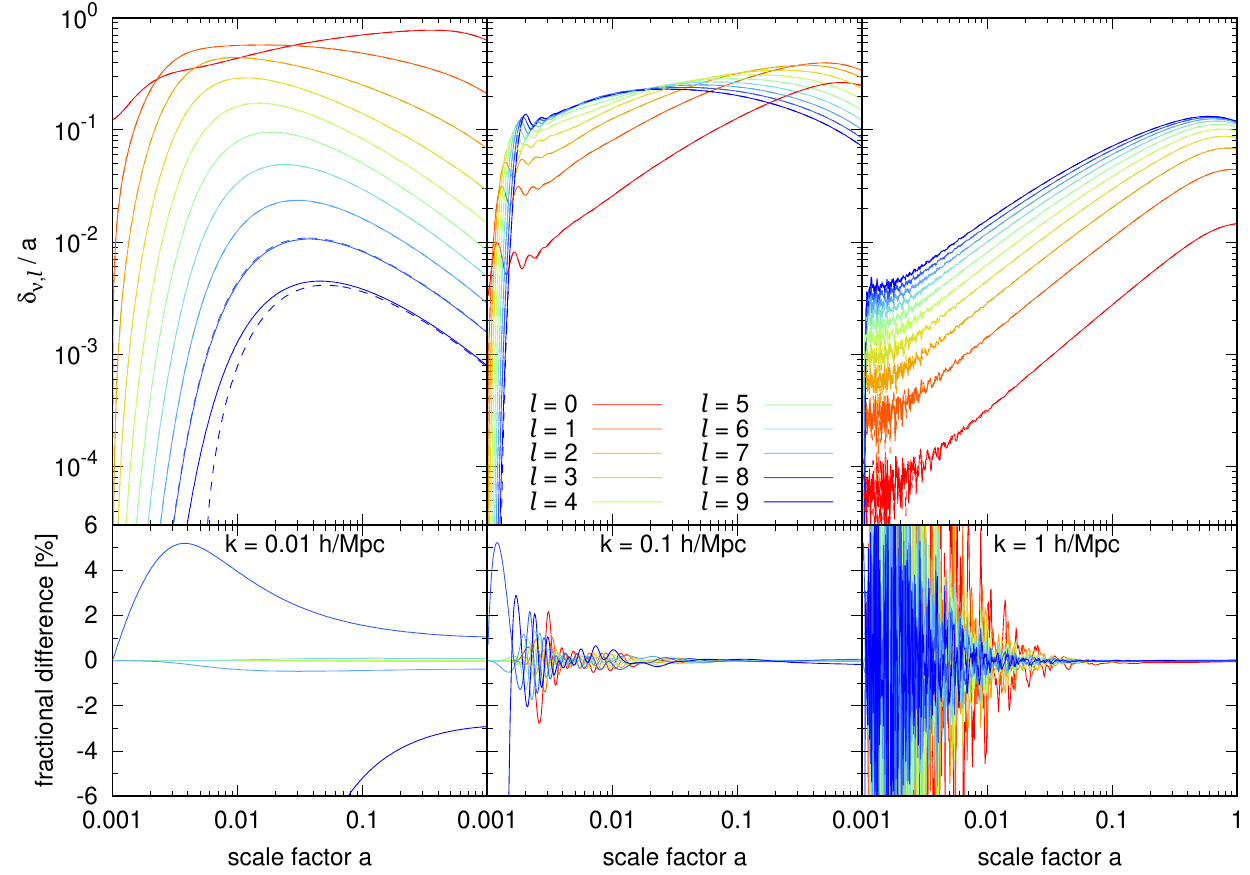}%
  \end{center}
  \caption{The first ten neutrino density Legendre coefficients
    $\delta_{\nu, \ell}$, $\ell=1, \ldots, 10$, averaged over  $N_\tau=10$ fluids,
    as functions of the scale factor~$a$ at different wave numbers. {\it Left}:  $k = 0.01~h/$Mpc.
   {\it Center}: $0.1~h$/Mpc.  {\it Right}: $1~h/$Mpc.	
    We consider $N_\mu=20$ (solid lines) or $N_\mu=10$ (dashed lines) for the
    $\nu\Lambda$CDM model of table~\ref{t:nuLCDM_reference}, which
    at $z=0$ has a neutrino free-streaming scale of $k \approx 0.2~h/$Mpc.
    The upper row shows the quantity $\delta_{\nu, \ell}/a$, while the lower row 
    plots the fractional difference in $\delta_{\nu, \ell}/a$ between the choices
    of $N_\mu=20$ and $N_\mu=10$.\label{f:Leg_coeff}
  }
\end{figure}

In equation~(\ref{e:d_truncation}) we truncated an infinite set of evolution equations for the Legendre coefficients of the neutrino density contrast at a finite number $N_\mu$ of coefficients.  Figure~\ref{f:Leg_coeff} tests this approximation.  For fixed $N_\tau=10$, the plots compare the first ten $\delta_{\nu,\ell} := \sum_{\alpha=0}^{N_\tau-1} \Omega_\alpha\delta_{\alpha,\ell}/\sum_{\alpha =0}^{N_\tau-1} \Omega_\alpha$ for $N_\mu=10$ and $20$.  If our ${\mathcal O}(N_\mu^{-2})$ truncation approximation is valid, then the two should agree at the percent level for the dominant $\ell$.  We compare them  in three regimes: $k=0.01~h/$Mpc (i.e., much smaller than the $z=0$ free-streaming scale $k_\mathrm{FS}\approx 0.2~h/$Mpc); $k = 0.1~h/$Mpc (i.e., $k \sim k_\mathrm{FS}$); and $k=1~h/$Mpc (i.e., $k \gg k_\mathrm{FS}$).

On large scales $k=0.01~h/{\rm Mpc} \ll k_{\rm FS}$, power flows from low to high $\ell$ slowly, leading $\delta_{\nu,\ell}$ to decline steeply with $\ell$
as shown in the top left panel of figure~\ref{f:Leg_coeff}.
  Then the truncation formula~(\ref{e:d_truncation}) is dominated by $\delta_{\alpha,\ellmax-2}$, causing the term proportional to $k v_\alpha / \Hc$ in equation~(\ref{e:dnu_eom_ell}) to have the wrong sign.  However, it is precisely in this $k \ll k_{\rm FS}$~regime that a large relative error in the largest multipole~$\ell$ does not affect the smaller $\ell$'s evolution.  Indeed, the bottom left panel of figure~\ref{f:Leg_coeff} confirms that
switching between the choices of $N_\mu=10$ and $20$ affects the $0 \leq \ell \leq 7$ terms only at the sub-percent level at all times.

On the other hand, the top middle panel shows that all  multipoles~$\ell$ are important for~$k \approx 0.2\,h/{\rm Mpc} \sim k_{\rm FS}$, 
as they generally all remain within an order of magnitude of one another.
 Observe also  that nonzero velocities~$v_\alpha$ lead to acoustic oscillations in the neutrino fluids at early times, $a < 0.01$.
  Since even a small phase difference between the  $N_\mu=0$ and $N_\mu=20$ calculations can lead to a large relative difference during oscillations, we find that the resulting $\delta_{\nu,\ell}$ can differ by several percent between choices of $N_\mu$ at $a < 0.01$.  After these oscillations have died down, however, the relative errors are indeed within $1\%$ at  $a \geq 0.01$.

At small scales $k=1~h/{\rm Mpc} \gg k_{\rm FS}$, we again observe acoustic oscillations in the top right panel of figure~\ref{f:Leg_coeff}, this 
time with even larger amplitudes and correspondingly larger relative differences between the $N_\mu=10$ and $N_\mu=20$ calculations.  Nevertheless, the dominant $\delta_{\nu,\ell}$ terms have converged to $<1\%$ by $a = 0.01$  and all others by $a = 0.02$.  In particular, $\delta_{\nu,9}$, whose evolution equation is directly modified by the truncation formula~(\ref{e:d_truncation}) for $N_\mu=10$ but not for $N_\mu=20$, converges to better than $0.1\%$ for $a \geq 0.04$.

\subsubsection{Varying $N_\tau$ and $N_\mu$}

\begin{figure}[t]
    \begin{center}
      \includegraphics[width=150mm]{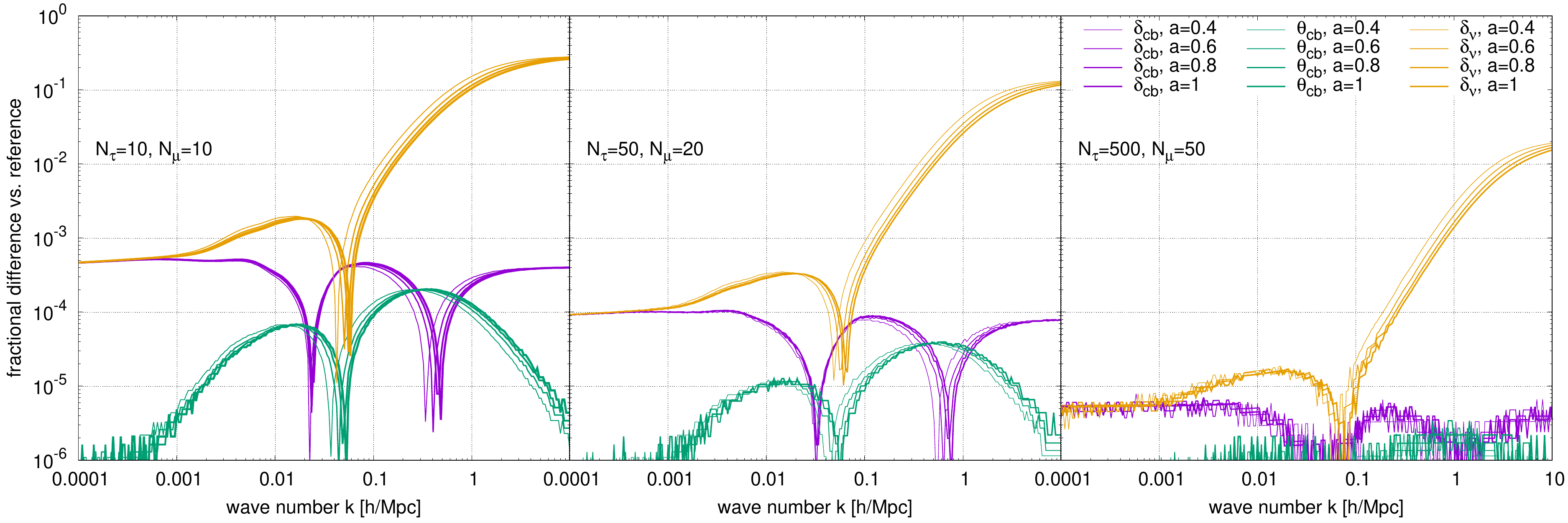}%
    \end{center}
    \caption{Comparison of multi-fluid predictions of the CDM+baryon density contrast $\delta_{\rm cb}$, velocity divergence $\theta_{\rm cb}$, and the monopole of the total neutrino density contrast $\delta_{\nu,0}$ at various scale factors~$a$ and wave number~$k$, 
      between different choices of $N_\tau$ and $N_\mu$  in the calculation.  {\it Left}: $N_\tau=10$ and $N_\mu=10$.  {\it Center}: $N_\tau=50$ and $N_\mu=20$. {\it Right}: $N_\tau=500$ and $N_\mu=50$.    All quantities have been plotted as a fractional difference from their high-accuracy reference values computed from a ($N_\tau=1000$, $N_\mu=100$)-run.\label{f:mflr_convergence}
  	}
  \end{figure}

Next, we show that our results converge as $N_\tau$ and $N_\mu$ are increased.  We are particularly interested in the convergence of the total neutrino density monopole $\delta_{\nu,0}$, which directly affects the gravitational potential~\eqref{e:Poisson}.
  Figure~\ref{f:mflr_convergence} compares the neutrino density monopoles as well as the CDM+baryon density and velocity perturbations computed for several ($N_\tau$, $N_\mu$)-combinations, to a high-accuracy reference calculation using $N_\tau=1000$ and $N_\mu=100$ in a range of wave numbers $k$ and scale factors $a\geq 0.4$.
  
In the left panel, we see that even for modest numbers of neutrino fluids, $N_\tau=10$, and angular polynomials, $N_\mu=10$, the late-time neutrino density monopole on scales $k \ll k_{\rm FS}$ converges at the percent level.  Since it is precisely this $k \ll k_{\rm FS}$ regime in which neutrinos cluster the most strongly and hence contribute most noticeably to the gravitational potential, together with the suppression factor $f_\nu = \Omno / \Ommo \lesssim 10\%$ in Poisson's equation~\eqref{e:Poisson}, 
we can conclude  on this basis that the choice of $N_\tau=N_\mu=10$ suffices for a $\sim 0.1\%$-level calculation of CDM+baryon clustering in the presence of massive neutrinos.

Of course, increasing $N_\tau$ and $N_\mu$ yields in principle even greater accuracy in the neutrino density.  Evidently in the middle panel of figure~\ref{f:mflr_convergence},  with the choice of $N_\tau=50$ and $N_\mu=20$ convergence improves to $< 0.1\%$ at $k\lesssim k_{\rm FS}$ and $<3\%$  at $k \lesssim 1~h/$Mpc.  Scales corresponding to $k \gtrsim 1\, h$/Mpc remain inaccurate because of acoustic oscillations.   However, this too can be controlled to the $2\%$ level up to  $k = 10~h/$Mpc if we substantially increase $N_\tau$ and $N_\mu$ to $500$ and $50$, respectively, as shown in the right panel of figure~\ref{f:mflr_convergence}.  The accuracies shown in the figure are consistent with the results of reference~\cite{Dupuy:2013jaa}, which at $k = 0.2~h/$Mpc finds a $0.1\%$ error for $N_\tau=40$ and a $0.01\%$ error for $N_\tau=100$.


\subsection{Comparison with standard perturbation theory}

\subsubsection{Comparison with integral linear response}
\label{subsubsec:ilr}

The method of integral linear response  refers to a formal solution of the linearised, non-relativistic Vlasov--Poisson system for the Fourier-space neutrino density contrast in the form~\cite{Bertschinger:1988ApJ...328...23B}:
\begin{equation}
\begin{aligned}
{\delta}_{\nu}(\varsigma,\vec{k}) \simeq \frac{3}{2} \Hc_0^2\, \Omega_{\rm m0} \int_{\varsigma_{\rm i}}^\varsigma \mathrm{d}\varsigma' \, a(\varsigma') \, \delta_{\rm m}(\varsigma',\vec{k}) \, (\varsigma-\varsigma') \, F\left[\frac{k(\varsigma-\varsigma')}{m_{\nu}}\right] \,  .
\label{e:ilr}
\end{aligned}
\end{equation}
Here, the time variable $\varsigma$ is the superconformal time defined via ${\rm d} \varsigma := {\rm d} \eta/a$, $F(q)$ is a scalar function that encapsulates the relativistic Fermi--Dirac distribution of the unperturbed neutrino population, and $\delta_{\rm m}$ is the formally external total matter density contrast to which the neutrino population gravitationally respond.

  Integral linear response has been previously applied to the investigation of neutrino clustering around cosmic string loops~\cite{Brandenberger:1987kf}, dark matter halos~\cite{Singh:2002de,Ringwald:2004np}  and more recently in $N$-body simulations of large-scale structure~\cite{AliHaimoud:2012vj,Chen2020a}.  Here, we contrast our multi-fluid approach with an explicit evaluation of the  integral linear response function~\eqref{e:ilr}.  Our implementation of the latter assumes a  source $\delta_{\rm m}$ history given at $a < \ain$ by the CDM+baryon growing mode~\eqref{e:cb_growing-mode-delta}  and  associated neutrino approximation~\eqref{e:dnu_tot_IC}, and at $a \geq \ain$ by the simultaneous solution of equations~\eqref{e:cdm+baryon-delta}, \eqref{e:cdm+baryon-theta}, and~\eqref{e:ilr}.  The  integration kernel is approximated with a fitting function following~\cite{AliHaimoud:2012vj}.

\begin{figure}[t]
	\begin{center}
		\includegraphics[width=75mm]{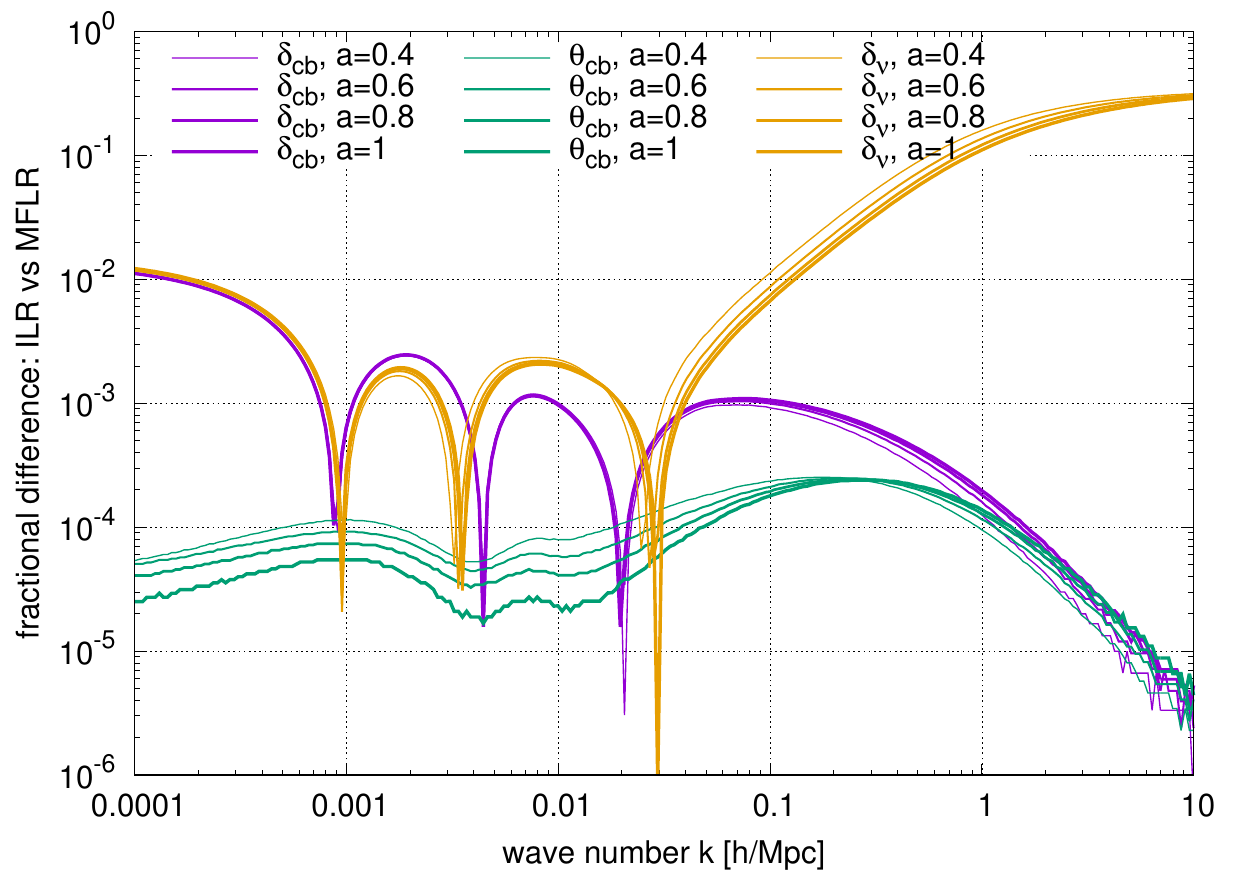}
		\includegraphics[width=75mm]{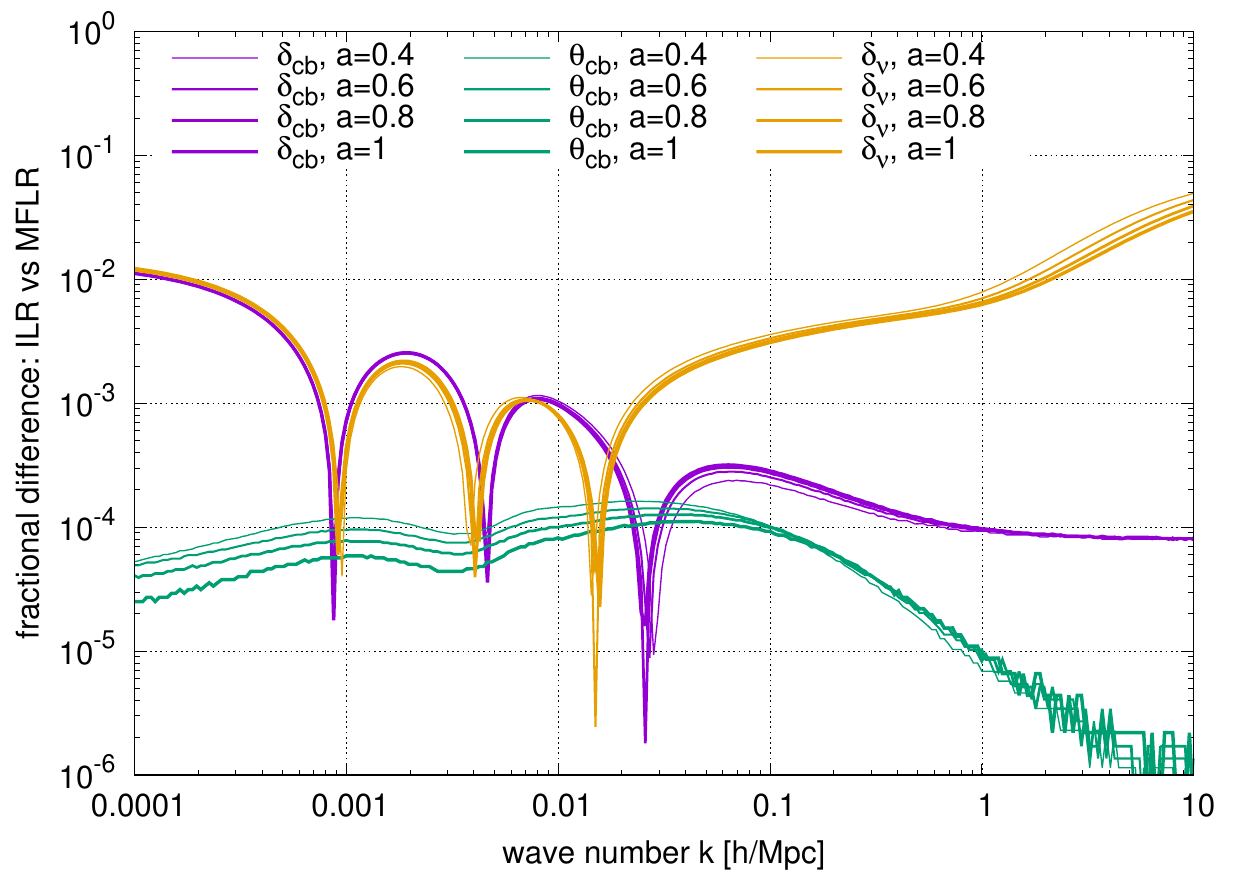}%
	\end{center}
	\caption{Fractional differences in the  predictions of $\delta_{\rm cb}$, $\theta_{\rm cb}$, and $\delta_{\nu,0}$ between the multi-fluid (MFLR) and the integral linear response (ILR) approach at various scale factors $a$, as functions of the wave number $k$.
{\it Left}: MFLR computations using $N_\tau=10$, a choice that is adequate for 0.1\%-accurate calculations of the CDM+baryon perturbations and 1\%-accurate predictions of the neutrino density contrast at $k \ll k_{\rm FS}$.  {\it Right}:   Choosing $N_\tau=1000$ improves the agreement in $\delta_{\nu,0}$ to 1\% up to $k \approx 1\, h$/Mpc.\label{f:mflr_ilr}
	}
\end{figure}

Figure~\ref{f:mflr_ilr} contrasts the multi-fluid and integral linear response  calculations.  We compare the neutrino density monopole $\delta_{\nu,0}$ summed over all $N_\tau$ neutrino fluids ---  identified with~$\delta_\nu$ in the integral linear response approach --- as well as the CDM+baryon density contrast~$\delta_\mathrm{cb}$ and velocity divergence~$\theta_\mathrm{cb}$ over a range of wave numbers at  $a \geq 0.4$.   Restricting our attention to $k \gtrsim 0.01~h/$Mpc, scales that are of greatest interest for data constraints and numerical simulations, the left panel of figure~\ref{f:mflr_ilr} shows once again that $N_\tau=N_\mu=10$ suffices to compute CDM+baryon perturbations at $0.1\%$ accuracy.  Furthermore, while the 1\%-level fractional differences at $k \lesssim 0.001~h/$Mpc between the two approaches are relatively large, they are nearly independent of the scale factor and particle type.  This means they can be reduced substantially by normalizing the matter power spectrum at $z=0$, as is conventionally done for the $\sigma_8$ cosmological parameter.  Put another way, the  growth factors and power spectra of the two methods at $a \gg \ain$ will agree to better than $1\%$.

Lastly, we note in the right panel of figure~\ref{f:mflr_ilr} that the choice of $N_\tau=1000$ and $N_\mu=100$ can bring the agreement between the multi-fluid and integral linear response methods to the sub-precent level even in the neutrino density contrast at $k \approx 1~h/$Mpc.   This setting may be useful if a high-accuracy computation of the neutrino density contrast is desired, e.g., for determining local clustering enhancement of the relic neutrino background.

\begin{figure}[t]
    \begin{center}
    \includegraphics[width=75mm]{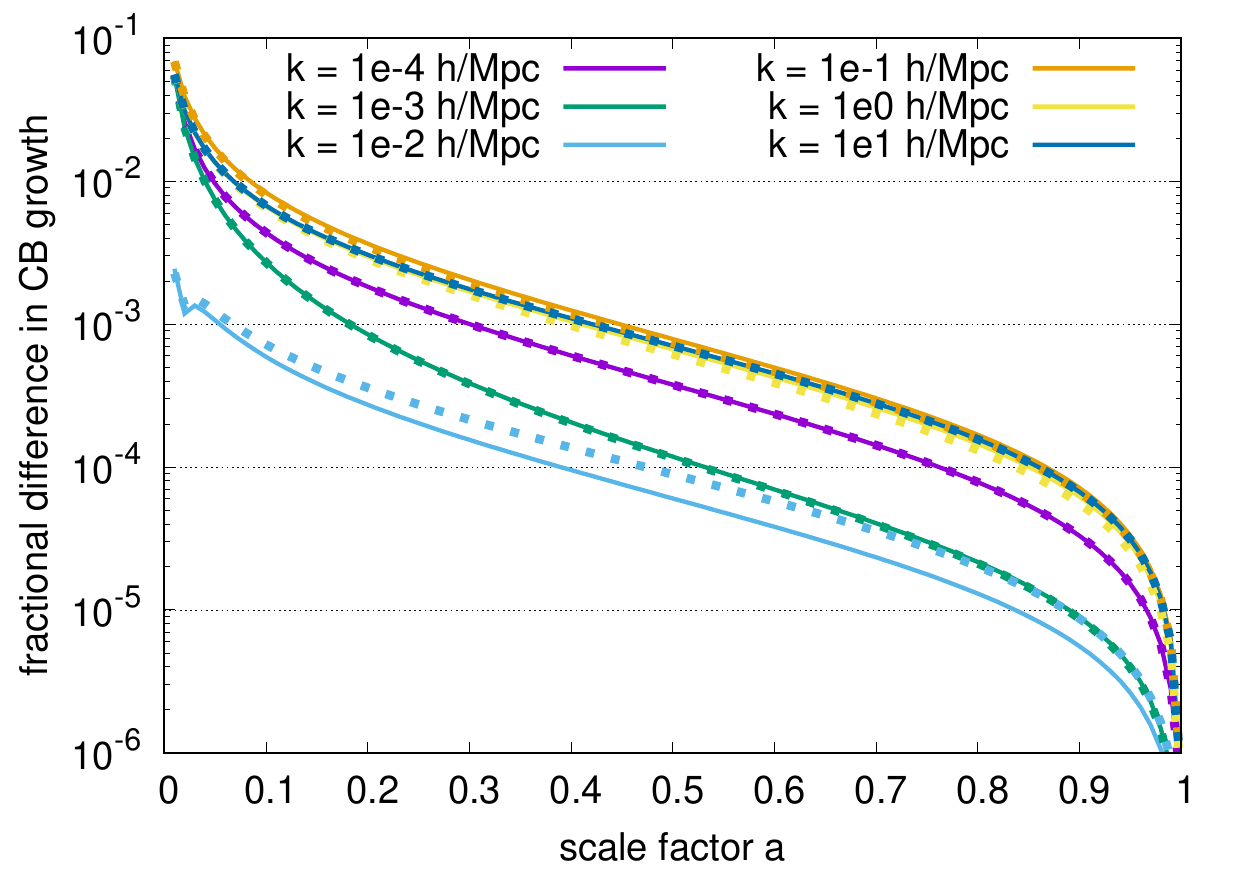}
    \includegraphics[width=75mm]{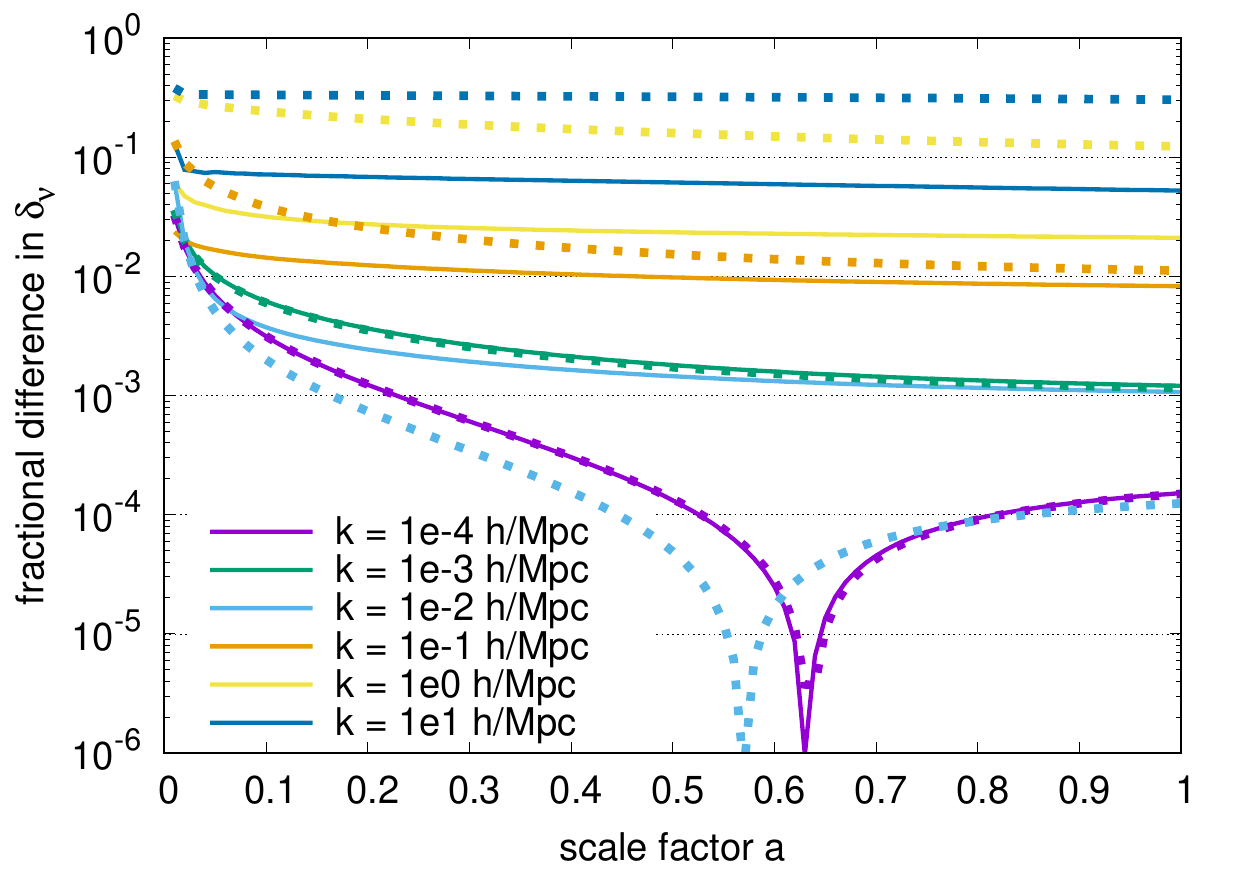}%
  \end{center}
  \caption{Fractional differences in the predictions between the multi-fluid approach and \classcode{}  as functions of the scale factor~$a$
  	for various wave numbers~$k$.
  	 {\it Left}: Differences in the normalised CDM+baryon growth factors. {\it Right}: Differences in the neutrino density contrasts~$\delta_{\nu,0}$. 
  	  Solid lines use $N_\tau=1000$ and $N_\mu=100$,
    while dotted lines represent $N_\tau=N_\mu=10$.
    \label{f:mflr_class}
  }
\end{figure}

\subsubsection{Comparison with \classcode{}}
The comparisons between the multi-fluid and integral linear response methods above are entirely based on our own numerical implemenations of the methods.
  Our next task, therefore, is to test our multi-fluid implementation against an independent linear Boltzmann  code \classcode{}~\cite{Lesgourgues:2011re},
  which computes the neutrino density contrast by solving the linearised (relativistic) Boltzmann equation decomposed into a Legendre hierarchy~\cite{Ma:1995ey}.  We run \classcode{} using the {\tt pk\_ref.pre} preset parameter tolerance file in order to obtain a high-accuracy matter power spectrum for comparison.

Figure~\ref{f:mflr_class} shows the fractional differences in the normalised CDM+baryon growth factor in the left panel and the neutrino density contrast in the right panel, both as functions of the scale factor~$a$, at different wave numbers~$k$ and for two choices of $(N_\tau, N_\mu)$.
  Both are consistent with our earlier results:  Firstly, even modest numbers of neutrino fluids and Legendre polynomials, $N_\tau=N_\mu=10$, suffice for a better-than-$0.1\%$-accurate calculation of the CDM+baryon growth on all scales at $a\gtrsim 0.5$;  percent-level inaccuracies at $a \lesssim 0.1$  are expected, considering that we treat CDM and baryons as a single fluid, and that our initial conditions as outlined in section~\ref{subsec:initial_conditions} do not precisely correspond to the growing-mode solution.  Secondly, percent-level errors in the late-time neutrino overdensity at $k \sim k_{\rm FS}$ for the choice of $N_\tau=N_\mu=10$, as well as more serious errors at $k \gg k_{\rm FS}$, can be mitigated by increasing $N_\tau$ and $N_\mu$.


\subsection{Summary}	

In summary, we have demonstrated the following about the multi-fluid linear treatment of neutrinos presented above.
\begin{itemize}
\item Late-time ($a \geq 0.4$) perturbations converge as the numbers of fluids $N_\tau$ and angular modes $N_\mu$ are increased, with the CDM+baryon density and velocity perturbations consistent at the $0.1\%$ level for $N_\tau = N_\mu = 10$.
\item Neutrino density perturbations agree at the percent level with predictions of the state-of-the-art integral linear response method~\cite{AliHaimoud:2012vj} up to $k \approx 0.1~h/$Mpc for $N_\tau = N_\mu = 10$, and up to $k \approx 2~h/$Mpc for $N_\tau=1000$, $N_\mu=100$.
\item The CDM+baryon growth factor normalised at $z=0$ agrees with the widely-used \classcode{} Boltzmann code at the $0.1\%$ level for $a \geq 0.4$.  Neutrino density perturbations agree to $\approx 2\%$ over that same range up to $k=0.1~h/$Mpc for $N_\tau = N_\mu = 10$, and up to $k=1~h/$Mpc for $N_\tau=1000$, $N_\mu=100$.
\end{itemize}
Thus, our perturbation theory is accurate to $0.1\%$ for CDM+baryons at all wavenumbers~$k$, and to $\approx 1\%$ for the neutrinos  below their free-streaming wave number~$k_{\rm FS}$, even for modest numbers of fluids and angular modes, $N_\tau = N_\mu = 10$.


\section{Neutrino linear response to non-linear clustering from Time-RG}
\label{sec:linear_response_to_non-linear_clustering}

Our ultimate goal in this article is to couple a multi-fluid linear treatment of neutrinos to a non-linear calculation of CDM+baryon clustering, in a linear response scheme akin to the integral linear response approach of reference~\cite{AliHaimoud:2012vj}.  In this section,  we take one more step in this direction by coupling the multi-fluid treatment to a non-linear perturbation theory for the CDM+baryons, namely, the Time-RG approach of references~\cite{Pietroni:2008jx,Lesgourgues:2009am,Audren:2011ne,Juergens:2012ap}.  The non-linear continuity and Euler equations of fluid dynamics form a hierarchy coupling each $N$-point correlation function to $(N+1)$-point functions, hence power spectra to bispectra and bispectra to trispectra.  Time-RG closes this hierarchy by neglecting the trispectra while evolving the power spectra and bispectra, with non-linear power spectra entering into mode-coupling integrals governing the evolution of each bispectrum.  This full Time-RG calculation can be sped up through a one-loop approximation that uses only the linear power spectra to evolve the bispectra.

Unlike standard perturbation theory, Time-RG by design can handle cosmologies in which the cold matter perturbations experience scale-dependent linear growth due to, e.g., neutrino free-streaming.  Our implementation is an adaptation of the {\tt redTime} code of reference~\cite{Upadhye:2015lia,Upadhye:2017hdl}, and employs the Fast Fourier Transform (FFT) acceleration techniques~\cite{Schmittfull:2016jsw,McEwen:2016fjn,Fang:2016wcf}.%
 \footnote{The FFT-accelerated version of {\tt redTime} is publicly available at {\tt github.com/upadhye/redTime}~.}
 For a broad overview of non-linear cosmological perturbation theories, see, e.g., references~\cite{Carlson:2009it,Bernardeau:2013oda}. Other higher-order perturbation theories capable of handling massive neutrino perturbations include~\cite{Saito:2008bp,Blas:2014hya,Fuhrer:2014zka,Levi:2016tlf,Aviles:2020cax,Garny:2020ilv}.

Our aim in this section is to estimate the enhancement to linear neutrino clustering when the CDM+baryon fluid is allowed to cluster non-linearly.  We quantify the portion of neutrinos clustering strongly enough to motivate a non-linear treatment, and we consider the impact of this enhanced neutrino clustering upon the power spectrum and bispectrum of the CDM+baryon fluid.


\subsection{Time-RG perturbation theory for CDM+baryons}

In order to model the time-dependent effects of neutrino clustering on non-linear CDM+ baryon perturbations, we use the full version of Time-RG, rather than the one-loop approximation applied in, e.g., the data analysis of~\cite{Upadhye:2017hdl}.  Full Time-RG and similar perturbation theories are known to suffer from a numerical instability at very small scales~\cite{Pueblas:2008uv,Valageas:2010rx,Vollmer:2014pma}, which drives the velocity auto-power spectrum to negative values.  References~\cite{Pueblas:2008uv,Valageas:2010rx,Vollmer:2014pma} trace this instability to the neglect of the velocity field's vorticity in the CDM+baryon equations of motion, thereby predicting that the instability should arise at length scales much smaller than the velocity dispersion length $\sigma_v(\eta) := [\int dk P_{\mathrm{cb}}(k) / (6\pi^2)]^{1/2}$, or, equivalently, $k \gg k_v := 2\pi / \sigma_v$.  We address this instability by smoothing the CDM+baryon perturbation inputs to the non-linear mode-coupling integrals of the Time-RG equations of motion, via the replacement $P_{\rm cb}(k) \rightarrow P_{\rm cb}(k) / (1 + k^2/k_v^2)$.  This smoothing stabilises Time-RG at $k \gg 1~h/$Mpc.

Observe also that Time-RG has been developed for Newtonian gravity~\cite{Pietroni:2008jx,Lesgourgues:2009am}.  Thus,to ensure that our computations agree with the outputs of such relativistic linear Boltzmann codes as \classcode{} and \cambcode{} on large scales, we use a ``scale back'' procedure when performing our non-linear perturbative computation:  Firstly, we switch off non-linear corrections altogether and run our Time-RG code to $z=0$ using the initial conditions of section~\ref{subsec:initial_conditions}.  This exercise yields  $\delta_\mathrm{cb}(k)$  and $\delta_{\alpha, \ell}(k)$ at $z=0$.  Secondly,
given the $z=0$ total matter power spectrum $P_\mathrm{m}(k)$ from \cambcode{} or \classcode{}, we define the $k$-dependent normalisation ${\mathcal N}(k)$ via
\begin{equation}
  \frac{{\mathcal N}(k)}{\Ommo}
  \left[
    \Omcbo \delta_\mathrm{cb}(k)
    + \frac{\Omno}{N_\tau} \sum_{\alpha=0}^{N_\tau-1} \delta_{\alpha,0}(k)
    \right]
  =
  \sqrt{P_\mathrm{m}(k)}\, .
  \label{eq:norm}
\end{equation}
Finally, we rerun the perturbation theory from the initial scale factor $\ain$, but this time with (i) the non-linear corrections restored, and (ii) the growing-mode initial perturbations of section~\ref{subsec:initial_conditions} all multiplied by ${\mathcal N}(k)$.  This procedure ensures that non-linear mode-couplings are computed using the properly-normalised linear power spectrum.

\begin{figure}[t]
	\begin{center}
		\includegraphics[width=75mm]{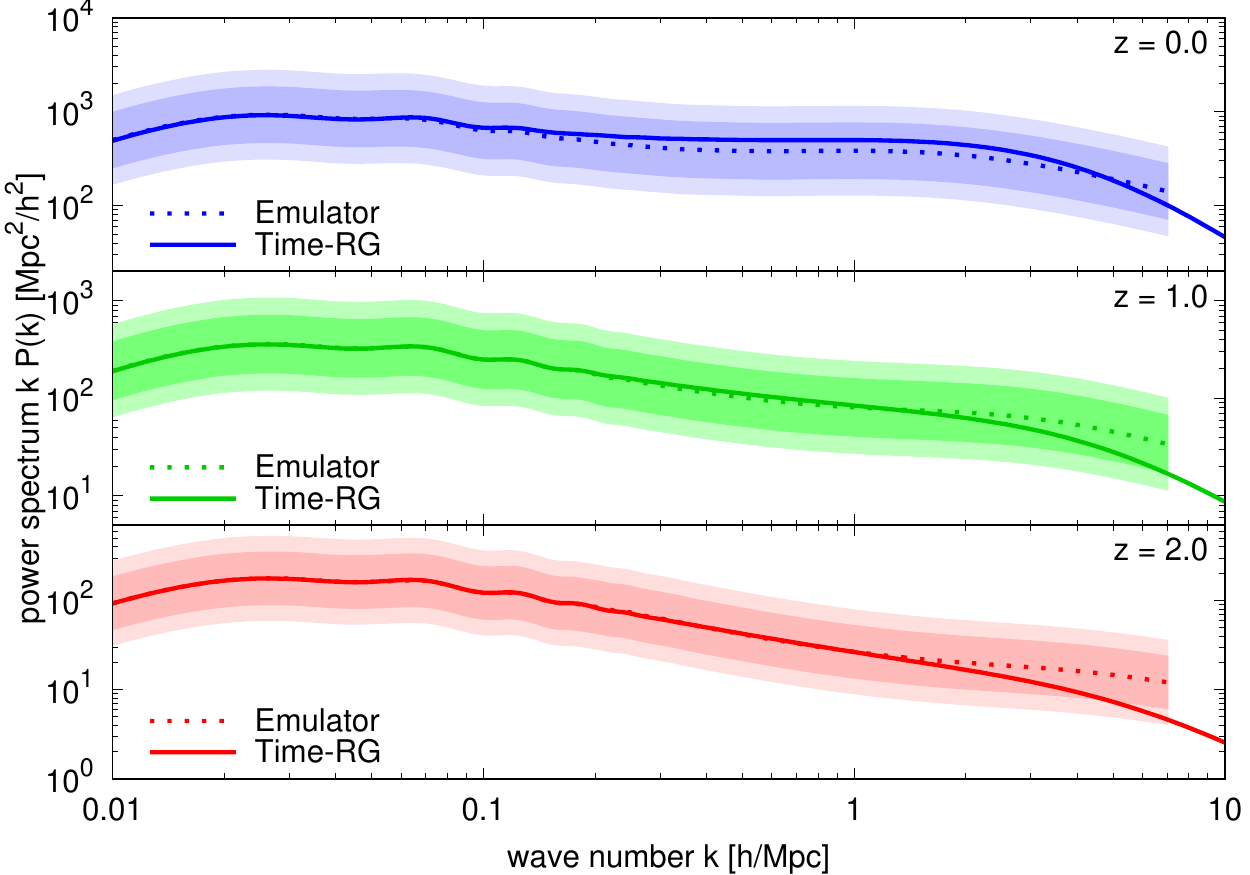}
		\includegraphics[width=75mm]{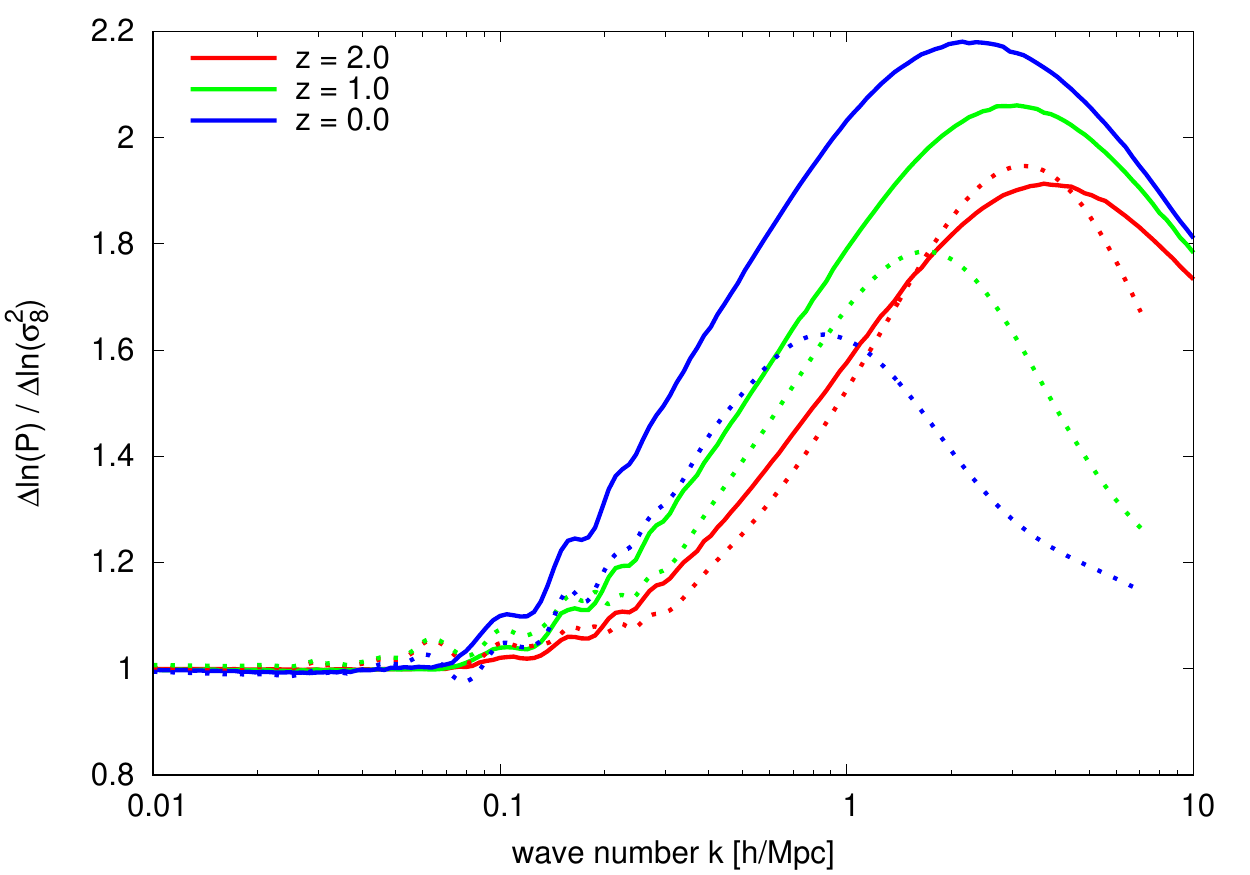}%
	\end{center}
	\caption{Comparison of our Time-RG implementation (solid lines) to the
		predictions of the {\tt CosmicEmu} $N$-body simulation emulator~\cite{Lawrence:2017ost} (dotted lines) at various redshifts~$z$ and wave numbers~$k$.
		{\it Left}: Predictions of the CDM+baryon power spectrum.   The inner and outer shaded regions correspond respectively to factors of $1/2$ to 2 and factors of $1/3$ to 3 of the emulator power.  {\it Right}: Logarithmic derivative of the CDM+baryon power spectrum with respect to the linear $\sigma_8^2$.  This derivative quantifies the sensitivity of the non-linear power to small changes  in the linear power spectrum normalisation.\label{f:trg_vs_emu}}
\end{figure}

Figure~\ref{f:trg_vs_emu} compares the CDM+baryon power spectrum outputs of our Time-RG implementation and of the {\tt CosmicEmu} $N$-body simulation emulator~\cite{Lawrence:2017ost} for the reference cosmology of table~\ref{t:nuLCDM_reference}.
 Though perturbation theory is not a precision tool at $k \gg 0.1~h/$Mpc, it agrees with the {\tt CosmicEmu} output up to a multiplicative factor of two or three over the entire $k$-range covered by the emulator, evident in the left panel of figure~\ref{f:trg_vs_emu}.

The right panel of figure~\ref{f:trg_vs_emu}, on the other hand, shows the logarithmic derivative of the non-linear CDM+baryon power spectrum with respect to the linear power spectrum normalisation $\sigma_8^2$.  This derivative approaches unity at large scales and rises at smaller scales as non-linearity enhances growth, and is a useful measure of the sensitivity of the non-linear power spectrum to small changes to the linear power spectrum normalisation such as represented by free-streaming neutrinos.
 Here, we see that while the Time-RG prediction of the logarithmic derivative agrees with that of {\tt CosmicEmu} at $z=2$, at later times the former overeshoots the latter by a factor of two.  We therefore conclude on the basis of figure~\ref{f:trg_vs_emu}  that Time-RG  suffices for estimating the impact of non-linear CDM+baryon clustering at the factor-of-two level.


\subsection{Clustering in the presence of CDM+baryon non-linearities}
\label{subsec:clustering_in_presence_of_cb_NL}

Equipped with the Time-RG theory for non-linear CDM+baryon perturbations, coupled to a multi-fluid linear theory for neutrinos that respond to the CDM+baryon non-linearities, we are now in a position to answer several important questions in an approximate way:
\begin{enumerate}
\item By how much is the linear clustering of neutrinos enhanced when
  the CDM+baryon fluid is allowed to cluster non-linearly?
\item What portion of the neutrino population clusters strongly enough 
  that linear response  is no longer adequate?
\item How much does neutrino clustering, as modelled by linear response
  calculations, enhance the clustering of the CDM+baryon fluid?
\end{enumerate}
These questions are addressed in three subsections below using our Time-RG plus multi-fluid perturbation theory.   We shall revisit them later on in section~\ref{sec:mflr_in_N-body_simulations} using more accurate calculations of non-linear CDM+baryon clustering from $N$-body simulations


\subsubsection{Neutrino clustering enhancement}

\begin{figure}[t]
	\begin{center}
		\includegraphics[width=130mm]{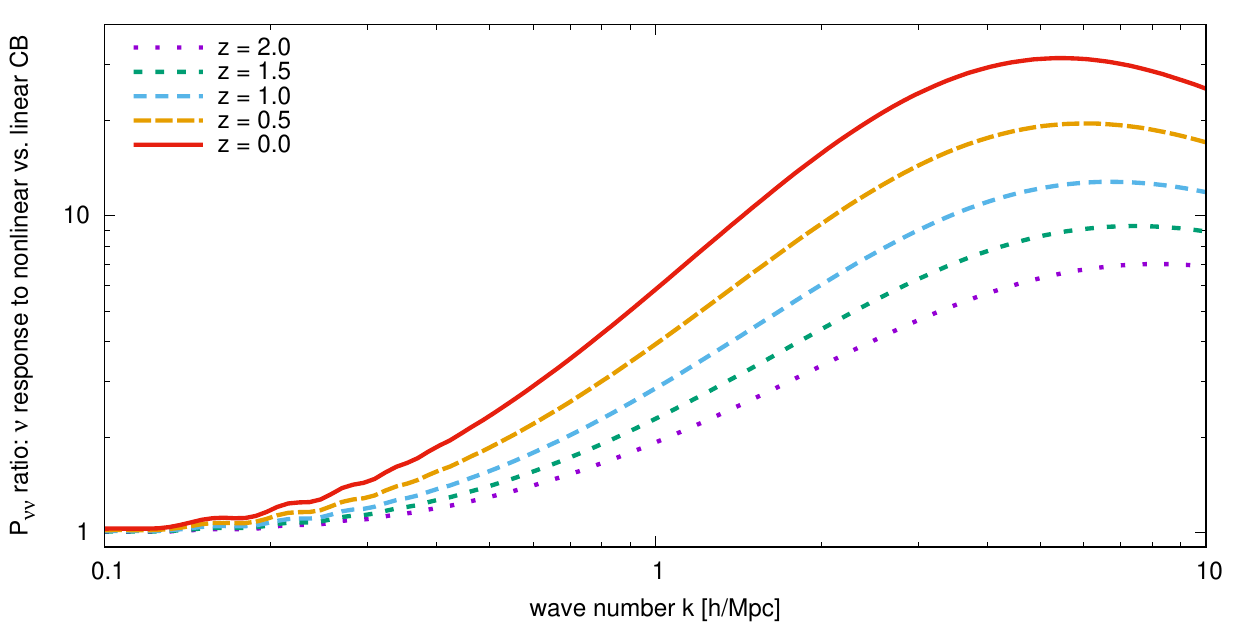}%
	\end{center}
	\caption{Ratios of the neutrino power spectra with and without neutrino linear response
		to non-linear CDM+baryon clustering at various redshifts.  Both sets of calculations use
		$N_\tau=200$ fluids and $N_\mu=20$ angular polynomials.
		\label{f:nu_enhancement}
	}
\end{figure}

A much-used approach to neutrinos in non-linear structure formation (e.g.,~\cite{Saito:2008bp,Agarwal:2010mt,Upadhye:2013ndm}) employs the linear neutrino perturbation output of \cambcode{} or \classcode{} as a source for non-linear CDM+baryon clustering.
 Because their clustering is determined in advance, neutrinos under this approach cannot respond  to non-linearities in the CDM+baryon sector.  In contrast, our Time-RG plus multi-fluid perturbation theory does allow neutrinos to respond linearly to non-linear CDM+baryon clustering; the amount of neutrino clustering is therefore generically expected to be enhanced relative to the purely linear case.

Figure~\ref{f:nu_enhancement} shows the ratio of neutrino power spectra with and without this enhancement.  At $z=0$, neutrinos at $k=1~h/$Mpc cluster $5.9$ times more strongly when they are allowed to respond to CDM+baryon non-linearity.  This enhancement factor rises to $31$ at $k = 5~h/$Mpc before gradually dropping again at higher wave numbers.  This peak wave number corresponds to galaxy cluster length scales, $2\pi / (5~h/{\rm Mpc}) \sim 1~h/$Mpc. Though perturbation theory is unreliable at these scales and does not track the formation of halos, our result is roughly consistent with that of reference~\cite{Ringwald:2004np},  which found an order-of-magnitude enhancement in the neutrino density at the centers of galaxy clusters.  Quantifying this enhancement is crucial for local measurements of the cosmic neutrino background such as by the proposed PTOLEMY experiment~\cite{Betti:2019ouf}.  Recent searches in references~\cite{Zhu:2013tma,Zhu:2014qma,Yu:2016yfe,Yu:2018llx} for qualitatively new cosmological neutrino effects will also be sensitive to neutrino clustering enhancements.


\subsubsection{Breakdown of neutrino linearity}
\label{subsubsec:breakdown_of_neutrino_linearity}

Since non-linear clustering of the CDM+baryon fluid can enhance neutrino clustering by over an order of magnitude, we must ask whether linear response remains reliable for neutrinos.  Given a power spectrum $P(k)$, the dimensionless power spectrum $\Delta^2(k) := k^3 P(k) / (2\pi^2)$ is a useful guide to the accuracy of linear perturbation theory.  We expect non-linear corrections to become important for $\Delta^2 \gtrsim 0.1$ and dominant for $\Delta^2 \gtrsim 1$.

\begin{figure}[tb]
	\begin{center}
		\includegraphics[width=75mm]{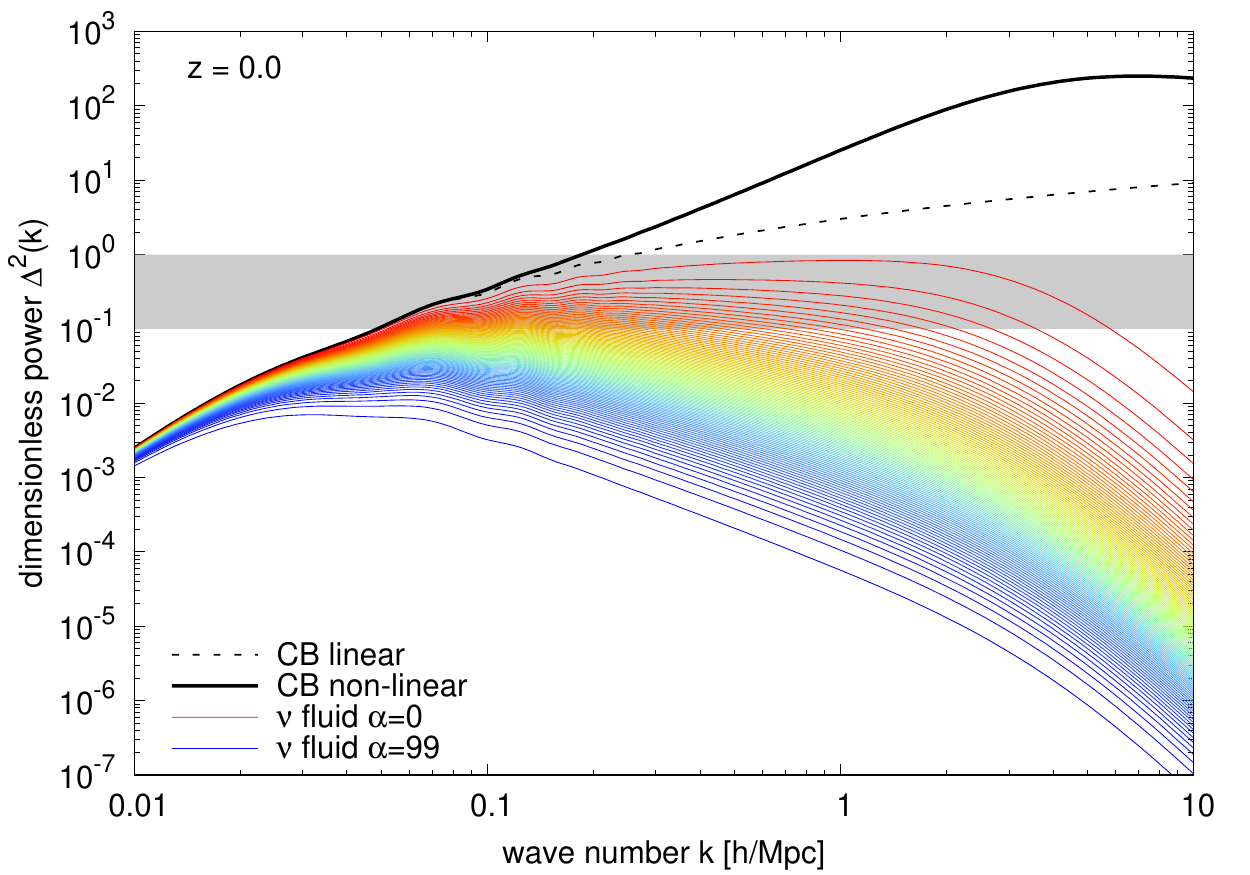}
		\includegraphics[width=75mm]{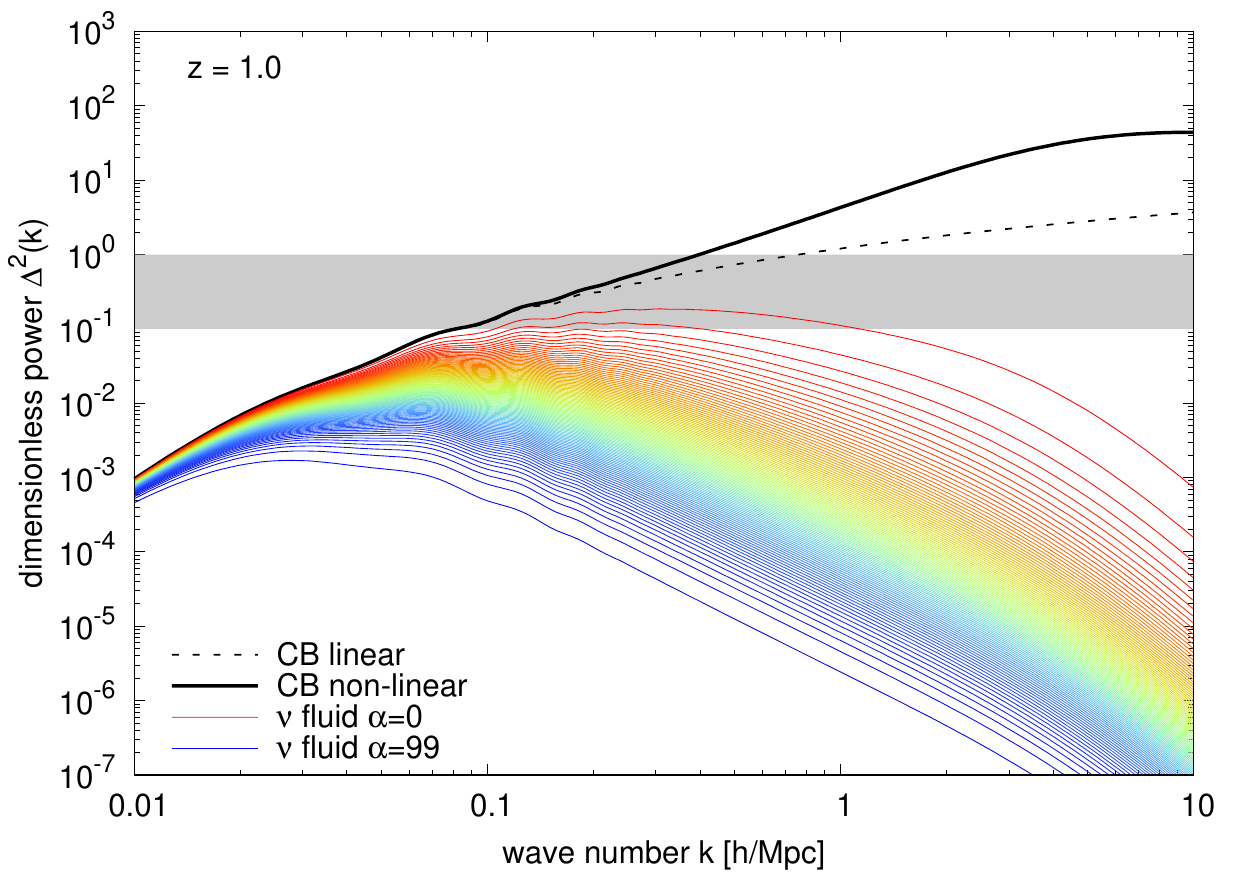}%
	\end{center}
	\caption{
		Dimensionless monopole power spectrum $\Delta^2(k) = k^3 P(k) / (2\pi^2)$
		for each of the $N_\tau=100$ neutrino fluids, with $N_\mu=20$, at two redshifts.
		{\it Left}: $z=0$.  {\it Right}: $z=1$.  
		Colors range from red for the lowest-momentum neutrinos to blue for the
		highest.  The solid black line denotes the non-linear CDM+baryon power
		spectrum, and the dashed black line its linear counterpart.  	
		The gray regions show the range $0.1 \leq \Delta^2 \leq 1$ in which
		linear perturbation theory breaks down.  For $\Delta^2 > 1$, even
		non-linear perturbation theory becomes increasingly unreliable.
		\label{f:D2_nu}
	}
\end{figure}

Figure~\ref{f:D2_nu} shows the dimensionless power spectra $\Delta^2_\alpha(k) := k^3 \delta_{\alpha,0}^2 / (2\pi^2)$ for each of $N_\tau=100$ neutrino fluids in the Time-RG+multi-fluid approach, as well as the corresponding non-linear CDM+baryon power spectrum.   For comparison, we also plot the CDM+baryon power spectrum computed from linear theory.
Judging from the difference between the linear (black dashed) and non-linear (black solid) CDM+baryon power spectra, our rule of thumb that non-linear corrections become important at $\Delta^2 \gtrsim 0.1$ appears valid.  By the time the linear CDM+baryon power spectrum reaches unity,  it underestimates its non-linear counterpart by about a factor of two.

\begin{figure}[t]
	\begin{center}
		\includegraphics[width=75mm]{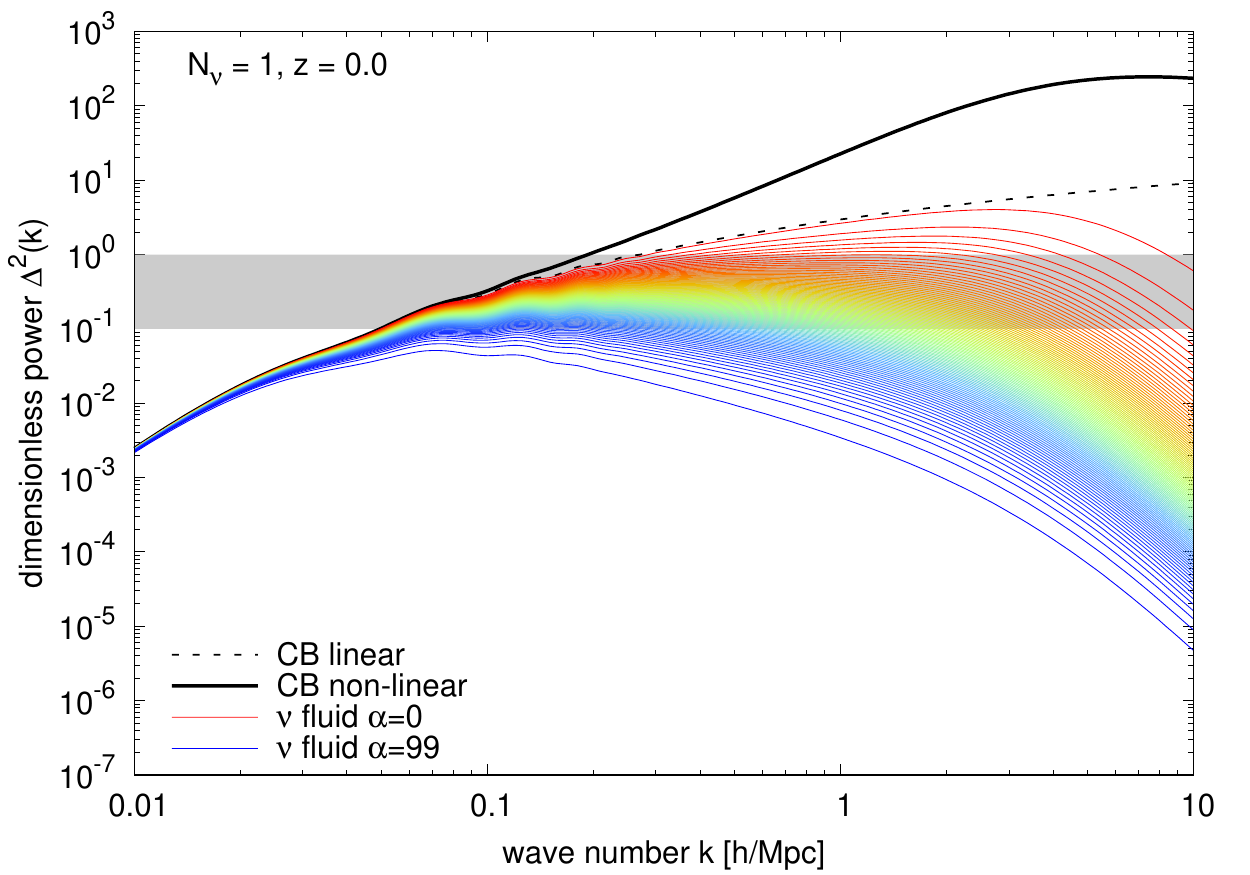}
		\includegraphics[width=75mm]{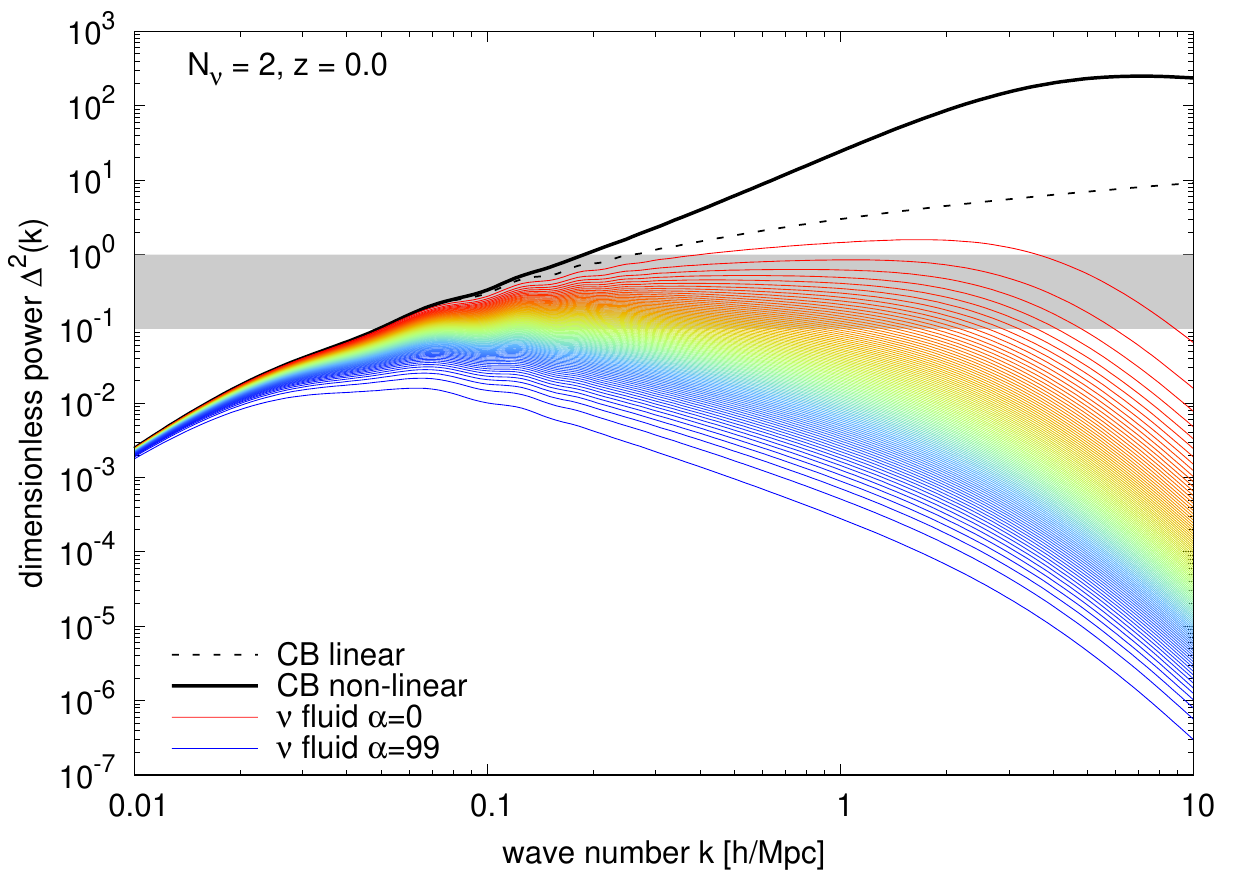}%
	\end{center}
	\caption{Same as the left panel of figure~\ref{f:D2_nu}, but for a cosmology in which the neutrino mass sum $\sum m_\nu=0.93$~eV is 
		distributed in different ways.  {\it Left}: Split between one massive and two massless species.  {\it Right}: Split between two massive (i.e., $m_\nu = 0.47$~eV each) and one massless species.  The other cosmological parameters are kept at their reference values reported in table~\ref{t:nuLCDM_reference}.
		\label{f:vary_Nnu}
	}
\end{figure}

In the case of the neutrino fluids, we find that the dimensionless power at $z=0$ exceeds $0.1$ for $27\%$ and $0.2$ for $8\%$ of the  fluids around the free-streaming scale $k \approx 0.2~h/$Mpc, indicating that linear response  is beginning to break down for a substantial portion of the total neutrino population.  In the most extreme case, power for the slowest neutrino fluid, $\alpha=0$, which represents $1\%$ of the population, peaks at $\Delta_0^2 = 0.83$ at $k \approx 1~h/$Mpc.  Even at $z=1$, the slowest $2\%$ of the neutrino fluids have $\Delta^2>0.1$.   Evidently, linear response theory is inadequate for a percent-level calculation of the neutrino power spectrum around the free-streaming scale.

Figure~\ref{f:vary_Nnu} explores the dependence of the neutrino power spectra on the number of massive neutrino species contributing to the total mass sum~$\sum m_\nu$.  This is of interest, because for the same $\sum m_\nu$, the effective free-streaming scale $k_{\rm FS}$ differs depending upon how exactly we distribute the masses amongst the neutrino families.
Recall that our reference model of table~\ref{t:nuLCDM_reference} and hence figure~\ref{f:D2_nu} assume three equal-mass neutrinos.  The left and right panels of figure~\ref{f:vary_Nnu} show cosmologies with the same $\Omno h^2 = 0.01$ and hence $\sum m_\nu = 0.93$~eV, but now with the neutrino mass sum distributed amongst one and two massive neutrino species respectively.   Because fewer massive species implies a greater mass for each one, the model with one single massive neutrino species clusters significantly more strongly than our reference model of figure~\ref{f:D2_nu}, with $\Delta^2$ exceeding $0.1$ and $1$ for $92\%$ and $6\%$ of the fluids respectively.  This is an interesting test case, as a single $\sim 1$~eV sterile neutrino consistent with the LSND/MiniBooNE anomaly~\cite{Aguilar-Arevalo:2018gpe} should exhibit similar clustering to that in the left panel of figure~\ref{f:vary_Nnu}.

\begin{figure}[t]
  \begin{center}
    \includegraphics[width=150mm]{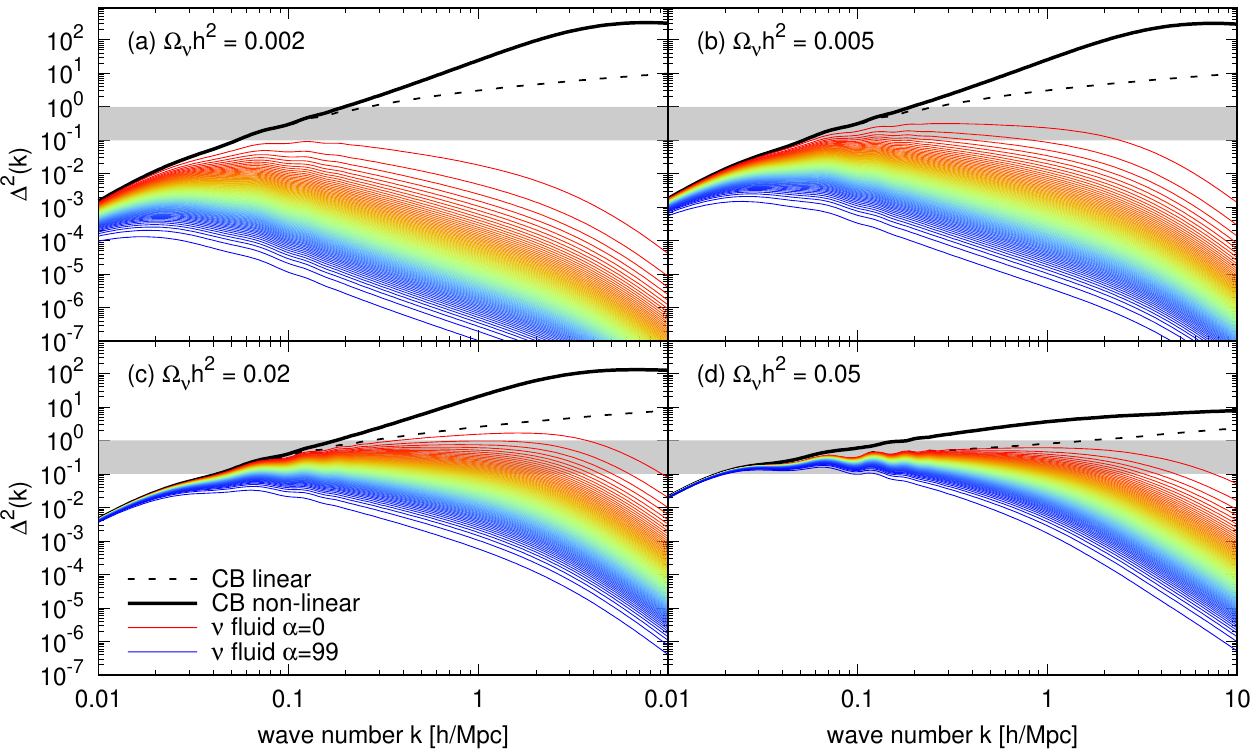}%
  \end{center}
  \caption{Same as the left panel of figure~\ref{f:D2_nu}, but for
    several different values of the neutrino energy density~$\Omno h^2$.
 {\it Top left}:   $\Omno h^2=0.002$.
   {\it Top right}: $\Omno h^2=0.005$.
    {\it Bottom left}: $\Omno h^2=0.02$.
    {\it Bottom right}:~$\Omno h^2=0.05$.  All other cosmological parameters
    are held fixed at their reference values reported in
    table~\ref{t:nuLCDM_reference}.  We assume three equal-mass neutrinos in
    each case.
    \label{f:vary_wnu}
  }
\end{figure}

Other variations away from our reference $\nu\Lambda$CDM model are considered in figure~\ref{f:vary_wnu}.  Here, we vary  the neutrino energy density in the range $\Omno h^2 \in [0.002, 0.05]$, while keeping the total matter density $\Ommo h^2$ and all  model parameters fixed at their reference values.  For the $\Omno h^2 = 0.002$ case (top left), the power of the slowest neutrino fluid peaks at $0.092$, meaning that this neutrino energy density is about the largest allowed if we wish to keep the fraction of neutrino fluids for which linear response might break down at below a percent.  The case of $\Omno h^2 = 0.005$ (top right) sits at just below the conservative bound of $\Omno h^2 < 0.006$~(95\%~C.I.)~\cite{Upadhye:2017hdl}; here, about $5\%$ of the neutrino fluids have power exceeding $\Delta^2=0.1$.  As  we increase the neutrino fraction to $\Omno h^2 = 0.02$ (bottom left), an even larger fraction,  about 75\%, of the neutrino fluids have entered the $\Delta^2>0.1$ regime.  In the extreme case of $\Omno h^2 = 0.05$ (bottom right), where massive neutrinos represent over a third of the total matter in the universe, all of the neutrino fluids exceed $\Delta^2=0.1$ in power.  Indeed, in the last case, the small-scale CDM+baryon clustering is so strongly suppressed by neutrino free-streaming that in order to maintain $\sigma_8$ at the reference value, we needed to increase the initial power spectrum normalisation by an order of magnitude.  These results are summarised in table~\ref{t:NL_fraction}, though we caution that they are model-dependent; the non-linear fraction would have to be recalculated for any substantial departure from our reference parameters of table~\ref{t:nuLCDM_reference}.

Lastly, we note that even though a non-linear neutrino calculation is beyond the scope of this article, our multi-fluid treatment points the way to such a calculation.  The multi-fluid perturbation theory is Lagrangian in momentum space.  This means that while the neutrinos' Eulerian momenta change as the system evolves, fluid elements cannot move from 
one fluid stream to another.  We are therefore free to  promote selectively an individual neutrino fluid's evolution from linear to non-linear.  This might be achieved, for example, by converting the slowest fluids into particles in an $N$-body simulation.  Existing ``hybrid'' simulations~\cite{Brandbyge:2009ce,Brandbyge:2018tvk,Bird:2018all} take a similar approach: In references~\cite{Brandbyge:2009ce,Brandbyge:2018tvk}, the entire neutrino population, initially tracked with a fully linear Boltzmann hierarchy, is converted to particles at the same time; Reference~\cite{Bird:2018all} divides the neutrino background into a high- and a low-momentum population, both tracked initially with linear response, with the latter converted to particles at some late time.  A multi-fluid treatment such as proposed in this work allows for a more fine-grained control of fluid-to-particle conversion, along with individual density and velocity power spectra for each of the fluids to be converted, as in figure~\ref{f:D2_nu}.

\begin{table}[tb]
	\begin{center}
		\footnotesize 
		\begin{tabular}{r|ccccc}
			\hline
			\hline
			$\Omno h^2$ &
			$0.002$ &
			$0.005$ &
			$0.01$ &
			$0.02$ &
			$0.05$
			\\
			
			Non-linear fraction &
			$<1\%$ &
			$5\%$ &
			$27\%$ &
			$75\%$ &
			$100\%$
			\\
			\hline
			\hline
		\end{tabular}
	\end{center}
	\caption{
		Fraction of neutrinos entering the non-linear regime at $z=0$, by the criterion
		$\Delta^2 > 0.1$.  All parameters other
		than $\Omno h^2$ are fixed at the values of our reference $\nu\Lambda$CDM
		model of table~\ref{t:nuLCDM_reference}.
		\label{t:NL_fraction}}
\end{table}



\subsubsection{CDM + Baryon clustering enhancement}

We have shown above in section~\ref{subsubsec:breakdown_of_neutrino_linearity} that neutrino linear response to non-linear CDM+baryon clustering can enhance the neutrino power spectrum by an order of magnitude or more.  This enhanced clustering will, in turn, source further non-linear CDM+baryon clustering, an effect that cannot be captured by power spectrum calculations without neutrino response.  Here, we demonstrate that this secondary enhancement of the CDM+baryon power is in fact very small, $< 0.5\%$ at late times, except in models with very large neutrino density fractions.

\begin{figure}[t]
	\begin{center}
		\includegraphics[width=75mm]{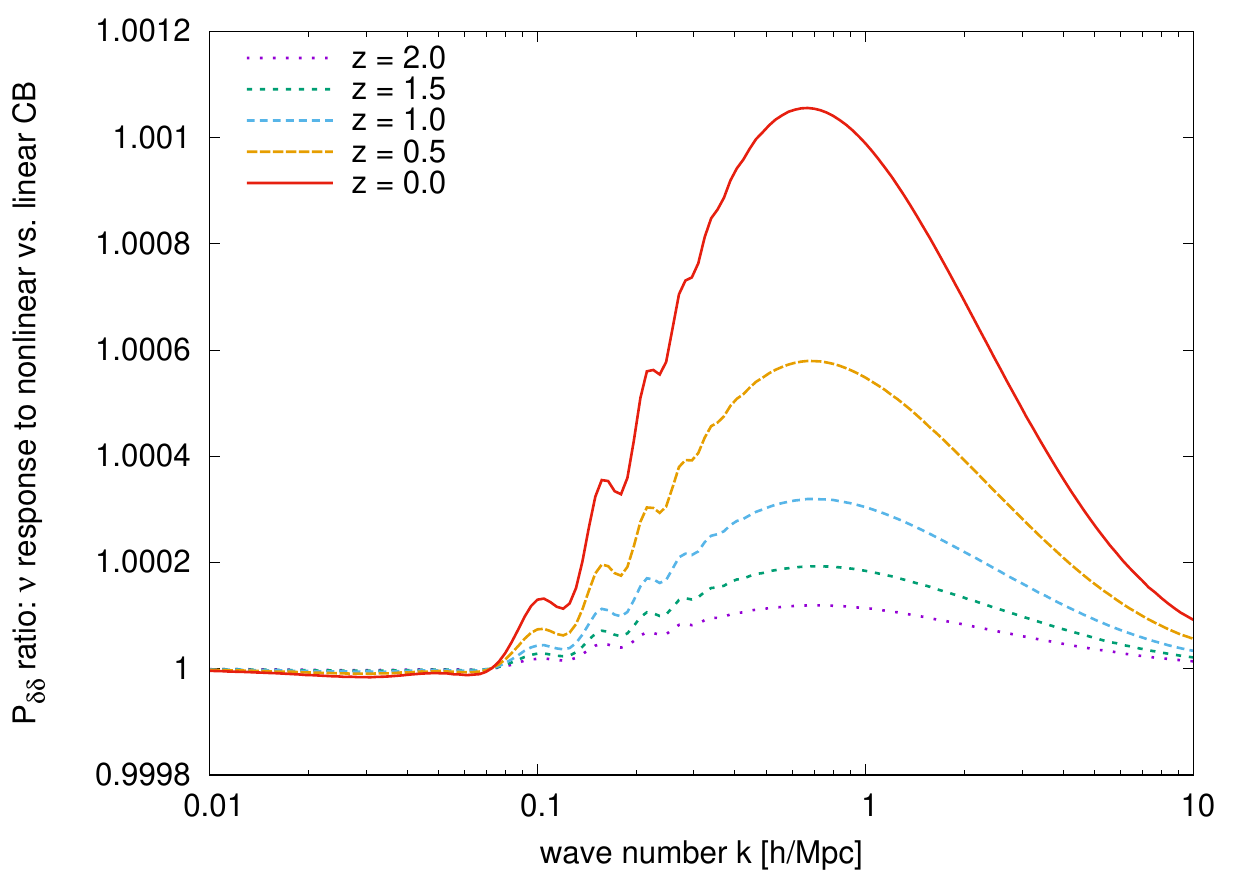}
		\includegraphics[width=75mm]{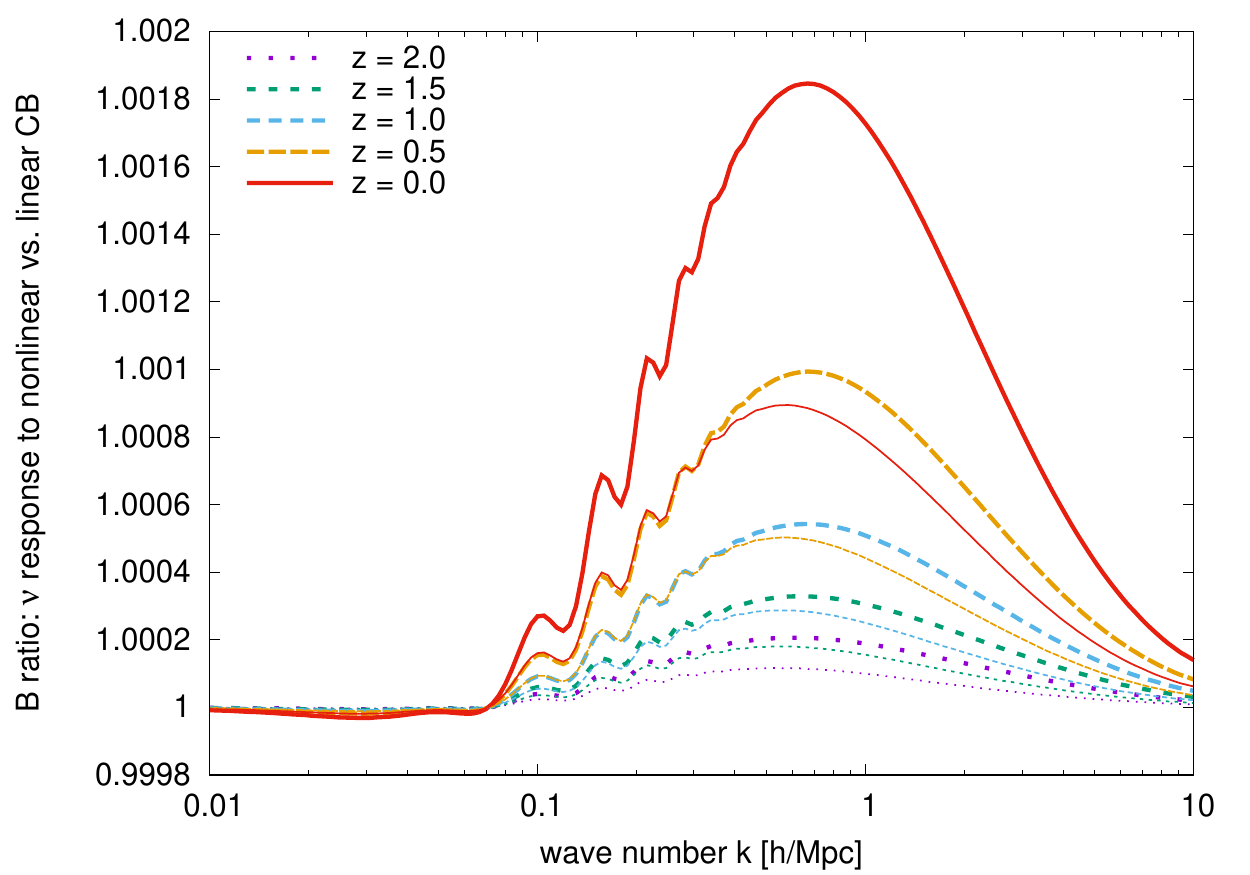}%
	\end{center}
	\caption{Ratios of CDM+baryon correlation statistics computed using Time-RG with and without neutrino linear response at various redshifts.
		{\it Left}: Power spectrum ratios.
		{\it Right}: Ratios of the equilateral bispectra (thick lines) and of the squeezed bispectra
		(thin lines).
		\label{f:cb_enhancement}
	}
\end{figure}

Figure~\ref{f:cb_enhancement} quantifies neutrino enhancement of CDM+baryon non-linear clustering in three ways: the power spectrum  in the left panel, and the equilateral and the squeezed bispectra in the right panel.  Time-RG is used to compute these CDM+baryon correlation statistics for the reference $\nu\Lambda$CDM model of table~\ref{t:nuLCDM_reference}, both with and without neutrino linear response.
Evidently, neutrino linear response enhances the CDM+baryon power spectrum by a maximum of $\approx 0.1\%$, occurring at $z=0$ around $k \approx 0.7~h/$Mpc. The squeezed bispectrum (thin lines) sees a similar fractional increase, and the equilateral bispectrum (thick lines) about twice the amount. 
This result is perhaps not surprising: although the $z=0$ neutrino power itself is enhanced by a factor of four at $k \approx 0.7~h/$Mpc according to figure~\ref{f:nu_enhancement}, we also see in the left panel of figure~\ref{f:D2_nu} that the bulk of the neutrino population still clusters at least two orders of magnitude less strongly than CDM+baryons on the same scale, even when neutrinos have been allowed to respond CDM+baryon non-linearity.  Coupled with a small neutrino fraction $\Omno / \Ommo = 7.5\%$, it is clear that the impact of enhanced neutrino clustering on the CDM+baryon perturbations cannot be but minute.

\begin{figure}[t]
	\begin{center}
		\includegraphics[width=50mm]{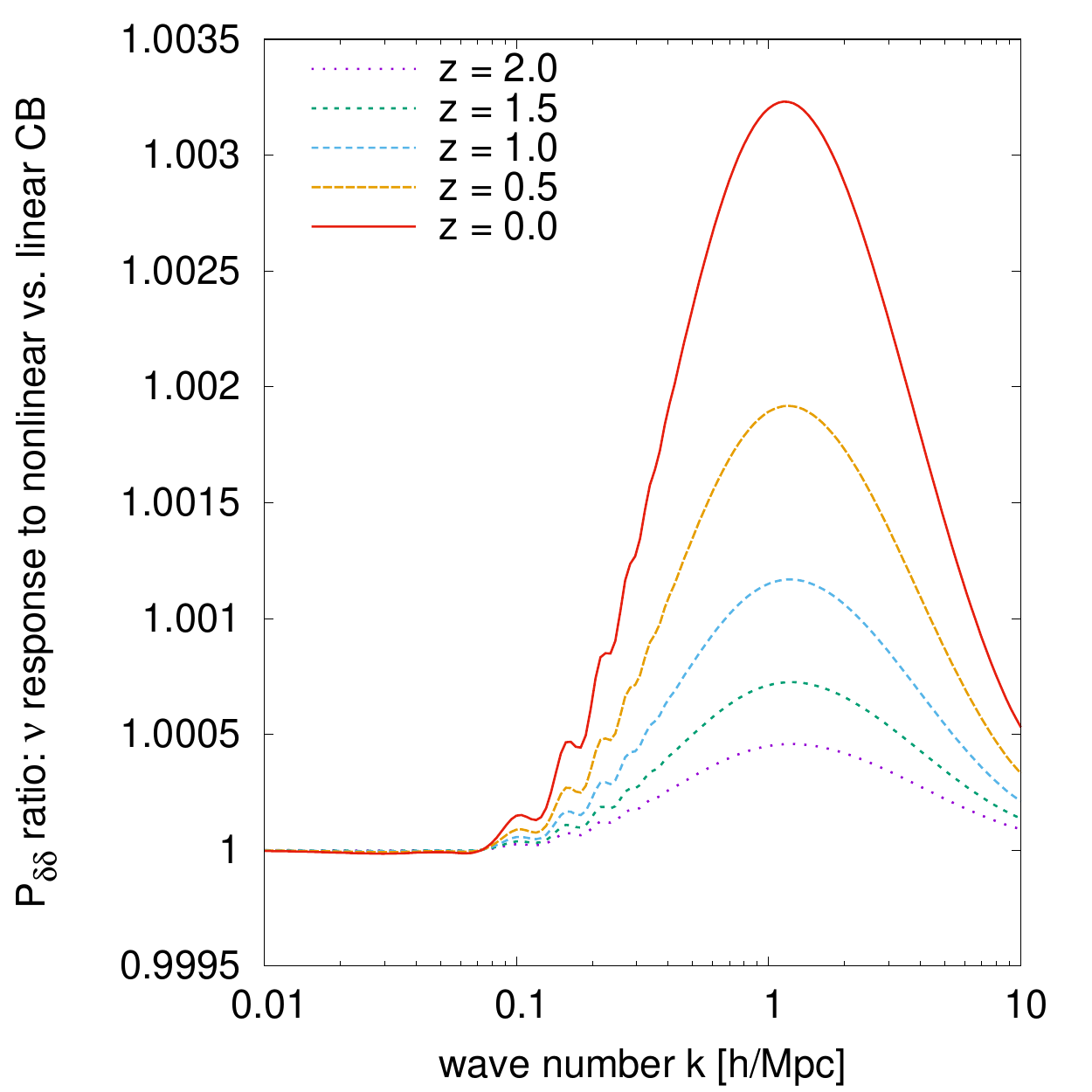}
		\includegraphics[width=50mm]{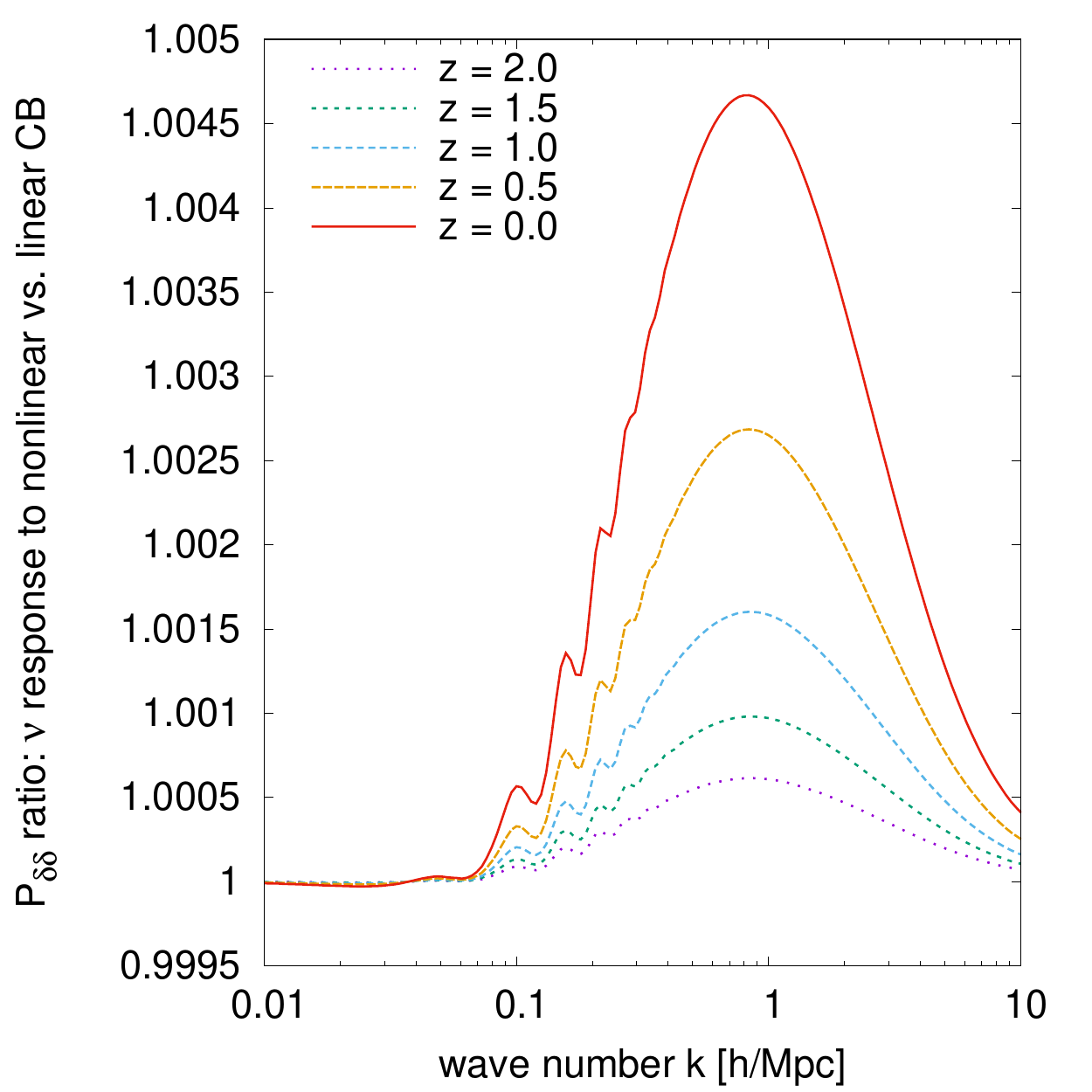}
		\includegraphics[width=50mm]{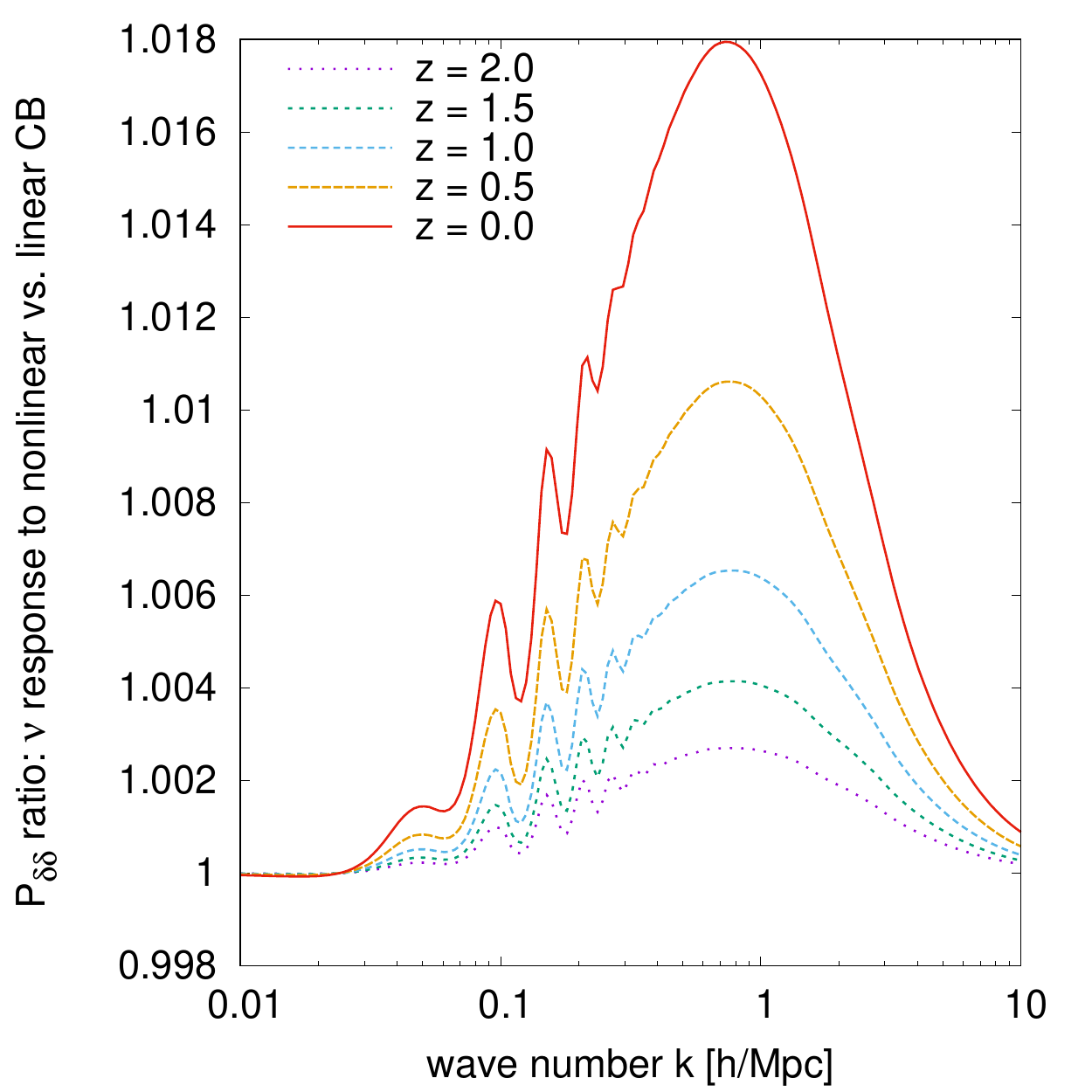}%
	\end{center}
	\caption{Same as the left panel of figure~\ref{f:cb_enhancement}, but for different assumptions about the neutrino sector.
		{\it Left}:~$\Omno h^2 = 0.01$ as in the reference $\nu\Lambda$CDM model, but now distributed in one massive and two massless neutrino species. {\it Center}:~$\Omno h^2 = 0.02$, distributed in three equal-mass species.   {\it Right}:~$\Omno h^2 = 0.05$, again for three equal-mass neutrinos.  All other cosmological parameters are fixed to their reference values displayed in table~\ref{t:nuLCDM_reference}.
		\label{f:cb_enhancement_variations}
	}
\end{figure}

It is possible to increase the CDM+baryon clustering enhancement by either concentrating all of the available neutrino energy density to one massive species (instead of spreading it across three), or simply increasing $\Omno h^2$ itself.  Figure~\ref{f:cb_enhancement_variations} explores both of these possibilities:  The left panel shows the reference case of  $\Omno h^2=0.01$ now distributed in one single massive and two massless neutrino species;  the center and right panels keep three equal-mass neutrino species but with $\Omno h^2$ increased to $0.02$ and $0.05$ respectively.  Evidently, while all three options augment the CDM+baryon enhancement somewhat, only in the $\Omno h^2 = 0.05$ case does the enhancement exceed $1\%$.

\subsection{Summary}

In summary, we have demonstrated in section~\ref{sec:linear_response_to_non-linear_clustering} that:
\begin{enumerate}
\item Time-RG perturbation theory for the CDM+baryon fluid is adequate for
  estimates of the non-linear power spectrum and its sensitivity to
  neutrino linear response at the factor-of-two level;
  
\item Neutrino linear response to non-linear CDM+baryon clustering leads to a factor-of-thirty increase in the neutrino power spectrum of our reference model  at wave numbers well beyond the free-streaming scale $k_{\rm FS}$, and 
is thus essential for  quantifying neutrino clustering on these scales;
 
\item For a wide range of $\Omno h^2$, including values consistent with current cosmological neutrino mass
  bounds, at least a few percent of the neutrino fluids have dimensionless
  power spectra $\Delta^2$ exceeding  $0.1$, which motivates a non-linear treatment of
  massive neutrinos if computing neutrino clustering acccurately is the end goal; and
  
\item The impact of enhanced neutrino clustering on the CDM+baryon fluid is however less than a percent
for a wide range of neutrino masses --- even some well beyond current cosmological limits --- and less than two
  percent for $\Omno h^2$ as high as $0.05$ (or, equivalently, $\sum m_\nu \approx 4.7$~eV).
\end{enumerate}
This concludes our exploration using Time-RG perturbation theory.  
In the next section, we shall implement our multi-fluid linear response description for neutrinos in an $N$-body simulation of the CDM+baryon component.  This will enable us to address more definitively the three questions posed in section~\ref{subsec:clustering_in_presence_of_cb_NL}.

\section{Multi-fluid neutrino linear response in $N$-body simulations}
\label{sec:mflr_in_N-body_simulations}

In this section we couple our multi-fluid neutrino linear response to an $N$-body description of CDM+baryon non-linear clustering.  For the latter, we use the publicly available \gadgetcode{} code~\cite{Springel:2005mi,Springel:2020plp}, a TreePM code that computes long-range forces by mapping particles to a mesh, making Fourier techniques applicable, and short-range forces via an oct-tree decomposition.  Time integration uses a variant of the leapfrog symplectic integrator.    We adopt the scale-back method to compute the matter power spectrum at the initialisation redshift and use  {\tt N-GenIC}~\cite{Angulo:2012ep} to set the initial conditions of the simulations in the Zel'dovich approximation.

We briefly describe first below how we modify \gadgetcode{} to work with our multi-fluid neutrino linear response theory,%
\footnote{Our modified code is available at {\tt github.com/joechenUNSW/gadget4-nu\_lr}~.}
before returning in section~\ref{subsec:testing_linear_response_in_simulations} to address the same three clustering questions considered in section~\ref{subsec:clustering_in_presence_of_cb_NL}.


\subsection{Modifications to Gadget-2}
\label{subsec:modification_of_gadget-2}

Massive neutrinos modify both the homogenous expansion rate of the universe and the inhomogeneous clustering of matter, both of which should be accounted for in our modification of \gadgetcode{}.   In the former case, neutrino kinetic energy leads in principle to a small deviation in the evolution of its energy density from the $a^{-3}$ behaviour.  In practice, however, this correction has no discernible effect on the  final outcome, as long as {\it the same} Hubble expansion rate is used in both the $N$-body code and in the scale-back initialisation procedure.

The latter, inhomogeneous effect constitutes the more important of the two classes of modifications, and we account for it in \gadgetcode{} via changes to its Particle--Mesh (PM) component.  Sepcifically, the gravitational force driving the inhomogeneous clustering of matter is computed in the PM component via the Fourier-space Poisson's equation~(\ref{e:Poisson}).   Because the stock version of \gadgetcode{} counts only contributions from the combined CDM+baryon fluid, i.e., only $\Omega_{\rm cb}(s)  \delta_\mathrm{cb}(s,k)$ appears inside the parentheses on the right-hand side of equation~(\ref{e:Poisson}), our first modification to \gadgetcode{} is  to multiply the default right-hand side by a factor
\begin{equation}
  F_\Phi
  :=
  1
  +
  \frac{\Omno}{\Omcbo N_\tau}
  \sum_{\alpha=0}^{N_\tau-1} \frac{\delta_{\alpha,\ell=0}(k)}{\delta_\mathrm{cb}(k)} .
  \label{e:f_Phi}
\end{equation}
As earlier, the index~$\alpha$ sums over the $N_\tau$ neutrino fluids, and the subscript $\ell=0$  identifies the monopole.  The density contrast ratios $\delta_{\alpha,0}(k) / \delta_\mathrm{cb}(k)$ are updated at every simulation time step using our multi-fluid linear response theory.

A step-by-step guide to how we run an CDM+baryon $N$-body simulation together with a multi-fluid update of the 
 density contrast ratios $\delta_{\alpha,0}(k) / \delta_\mathrm{cb}(k)$ is as follows:
\begin{enumerate}
\item Following equation~\eqref{eq:norm}, the density contrasts~$\delta_\mathrm{cb}$ and $\delta_{\alpha,0}$
  are normalised such that their density-weighted sum  
  \mbox{$(\Omcbo/\Ommo) \delta_\mathrm{cb}
  + (\Omno/\Ommo)\sum_{\alpha=0}^{N_\tau-1} \delta_{\alpha,0} / N_\tau$}
  evolved linearly to $z=0$ is equal to the square root of the desired $z=0$ total linear
  matter power spectrum as output by a linear Boltzmann code such as \cambcode{} or \classcode{}.   
  
\item Beginning at $\ain=10^{-3}$, we set the appropriately-normalised  initial conditions for $\delta_\mathrm{cb}$, $\theta_{\rm cb}$, and $\delta_{\alpha,\ell}$ according to section~\ref{subsec:initial_conditions}, and evolve them  from $\ain$ to the simulation start time $a_{\rm sim}=0.02$ using the linearised equations~\eqref{e:dnu_eom_ell} and \eqref{e:tnu_eom_ell} for the neutrinos, and 
equations~\eqref{e:cdm+baryon-delta} and~\eqref{e:cdm+baryon-theta} for the CDM+baryons.  The output $\delta_{\rm cb}$ and $\theta_{\rm cb}$ at $a_{\rm sim}=0.02$ set the CDM+baryon initial conditions for the $N$-body simulation --- the initialisation procedure is discussed in the companion paper~\cite{Chen2020a}.

\item \label{point3} 
From $a_{\rm sim}=0.02$ onwards, the true, non-linear evolution of the CDM+baryon perturbations is taken over by \gadgetcode{},
although we continue to solve the  linearised multi-fluid equations~\eqref{e:dnu_eom_ell}, \eqref{e:tnu_eom_ell}, \eqref{e:cdm+baryon-delta}, and~\eqref{e:cdm+baryon-theta} alongside.
At each simulation time step, \gadgetcode{} outputs a non-linear CDM+baryon density contrast~$\delta_\mathrm{cb}(k)$ averaged over all directions $\hat{k}$ for the same $k$.  This density contrast $\delta_\mathrm{cb}(k)$ is used to adjust its counterpart appearing in the linearised 
multi-fluid perturbative equations.  We likewise adjust the perturbative $\theta_\mathrm{cb}(k)$ by a factor $\delta_{\rm cb}^{\rm sim}(k)/\delta_{\rm cb}^{\rm pert}(k)$ formed from the density contrast outputs of the simulations and perturbative calculation at the simulation time step concerned.  These adjustments therefore source the neutrino linear response part of our calculation.

\item Between simulation time steps, the CDM+baryon and neutrino density and velocity perturbations are evolved using the linearised multi-fluid equations of motion.  Though our procedure leads to finite jump discontinuities in the perturbative $\delta_{\rm cb}$ and hence $\Phi$ at each simulation time step, integration of the neutrino equations of motion~\eqref{e:dnu_eom_ell} and~\eqref{e:tnu_eom_ell} results in neutrino perturbations that are in practice continuous in time.
  
\item The ratio $\delta_{\alpha,0} / \delta_\mathrm{cb}$ that appears in equation~\eqref{e:f_Phi} is formed 
at each simulation time step from the multi-fluid perturbative calculation, {\it before} we implement the  linear response adjustments outlined in step~\ref{point3}. 
\end{enumerate}

Following the above procedure, we have conducted two sets of simulations for the reference $\nu \Lambda$CDM cosmology of table~\ref{t:nuLCDM_reference}: a large-box set, in a cubic volume with edge length $L=1024$~Mpc$/h$; and a small-box set with $L=512$~Mpc$/h$.  Each set contains runs with $N=128^3, 256^3, 512^3$, and $1024^3$ particles.  The multi-fluid component always uses 
$N_\tau=50$ fluids and $N_\mu=20$ angular polynomials to represent the neutrino population.


\subsection{Convergence tests and comparisons with other approaches}
\label{subsec:testing_linear_response_in_simulations}

\begin{figure}[t]
  \begin{center}
    \includegraphics[width=75mm,height=65mm]{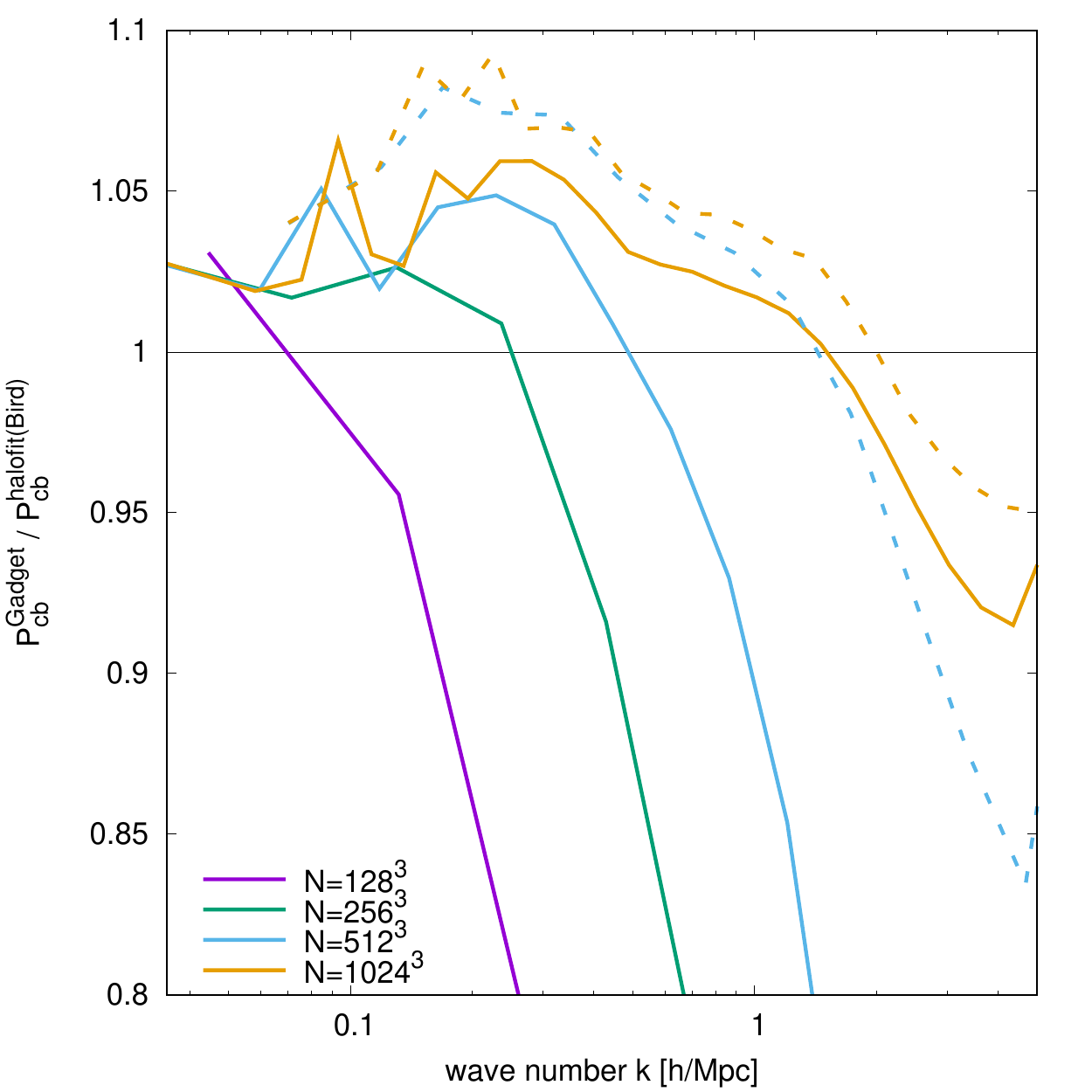}
    \includegraphics[width=75mm,height=65mm]{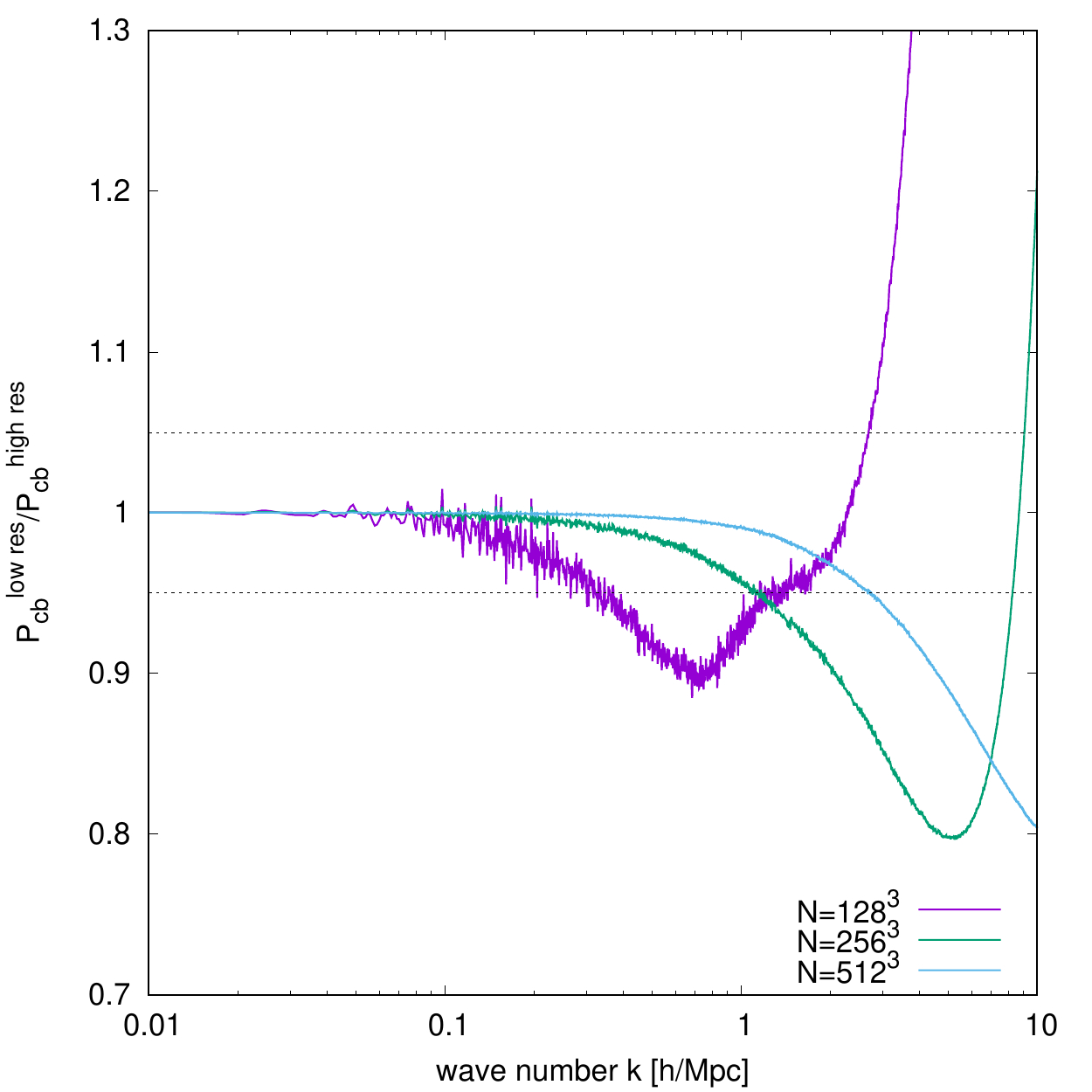}%
    
    \includegraphics[width=150mm]{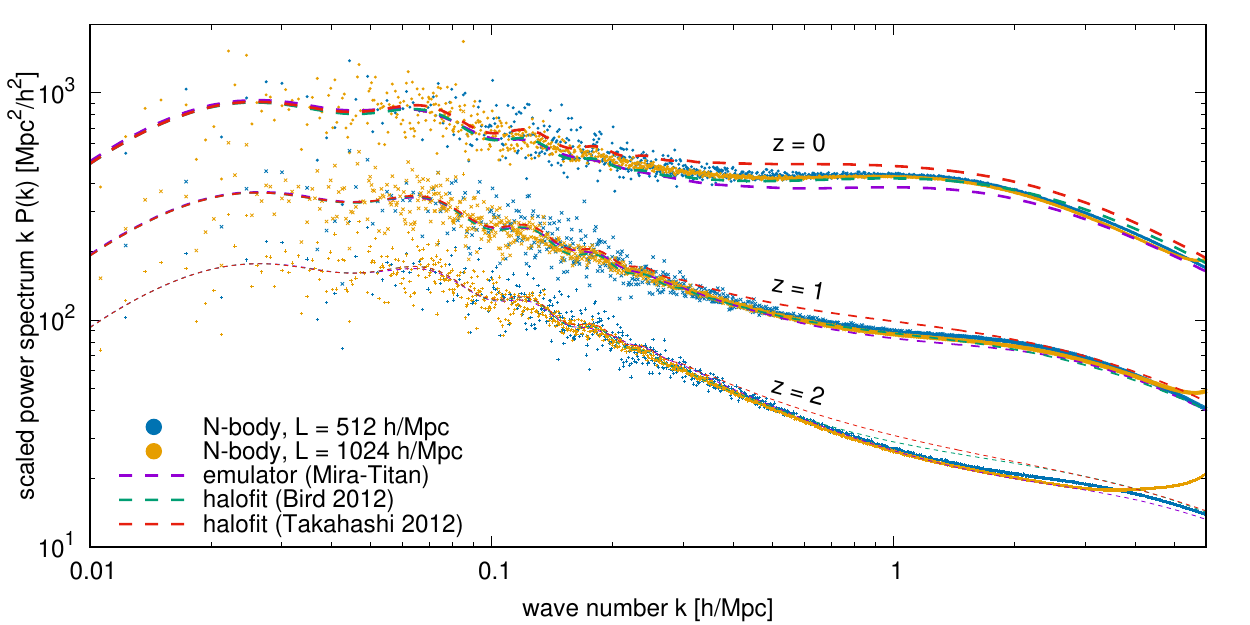}%
  \end{center}
  \caption{
    Convergence tests of the CDM+baryon power spectra from our \gadgetcode{}
    $N$-body simulations for the reference $\nu\Lambda$CDM model of
    table~\ref{t:nuLCDM_reference}.
    {\it Top left}:~Binned $z=0$ power spectra computed with various particle
    numbers~$N$ and in boxes of side lengths
    $L=1024$~Mpc$/h$ (large box; solid lines)
    and $L=512$~Mpc$/h$ (small box; dashed lines),
    normalised to the output of Halofit~\cite{Bird:2011rb}.
    {\it Top right}:~Ratios of the $N=128^3$, $256^3$, and $512^3$ small-box
    $z=0$ power spectra to the high-resolution 
    $N=1024^3$ result. To minimise sampling noise, these simulations have been
    initialised with the same random seeds.		
    {\it Bottom}:~Comparison of our $N=1024^3$ simulation results
    with the output of {\tt CosmicEmu}~\cite{Lawrence:2017ost}, 
    as well as the Halofit power spectrum fitting functions of
    references~\cite{Bird:2011rb} (Bird 2012) and~\cite{Takahashi:2012em}
    (Takahashi 2012) as implemented in \cambcode{}, at several redshifts.
    \label{f:nbody_emu_halofit_comparison}
  }
\end{figure}

Our convergence tests of this modified \gadgetcode{} code are shown in figure~\ref{f:nbody_emu_halofit_comparison}.  The top left panel shows our binned CDM+baryon power spectra normalised to the {\tt Halofit} power spectrum of reference~\cite{Bird:2011rb}, while the top right panel shows exclusively the power spectra from three small-box runs with $N=128^3$, $256^3$, and $512^3$, normalised to the high-resolution $N=1024^3$ output.  Evidently, the small-box  simulations are convergent at the $5$\% level up to $k=2~h/$Mpc.  

Measuring simulation accuracy, rather than precision, is however complicated by the $\sim 10\%$ disagreement between different state-of-the-art calculations in the literature, as shown in the bottom panel of figure~\ref{f:nbody_emu_halofit_comparison}.  Nevertheless, our large-box simulations agree with the output of {\tt  Halofit} 
 at the $5\%$ level for a broad range of wave numbers around $k_\mathrm{FS}\approx 0.2~h/$Mpc.  The $N=1024^3$ simulation in particular  finds 5\% agreement with both the {\tt Halofit} output at small wave numbers and the small-box results up to $k\approx 2~h/$Mpc.  A general $\sim 10\%$ agreement with other emulation and halo-fitting methods is also possible up to $k\approx 2~h/$Mpc.   Thus, the $L=1024$~Mpc$/h$, $N=1024^3$ simulation setting appears a suitable choice for the investigation of neutrino effects in the vicinity of the neutrino free-streaming scale.
 
\begin{figure}[t]
	\begin{center}
		\includegraphics[width=75mm]{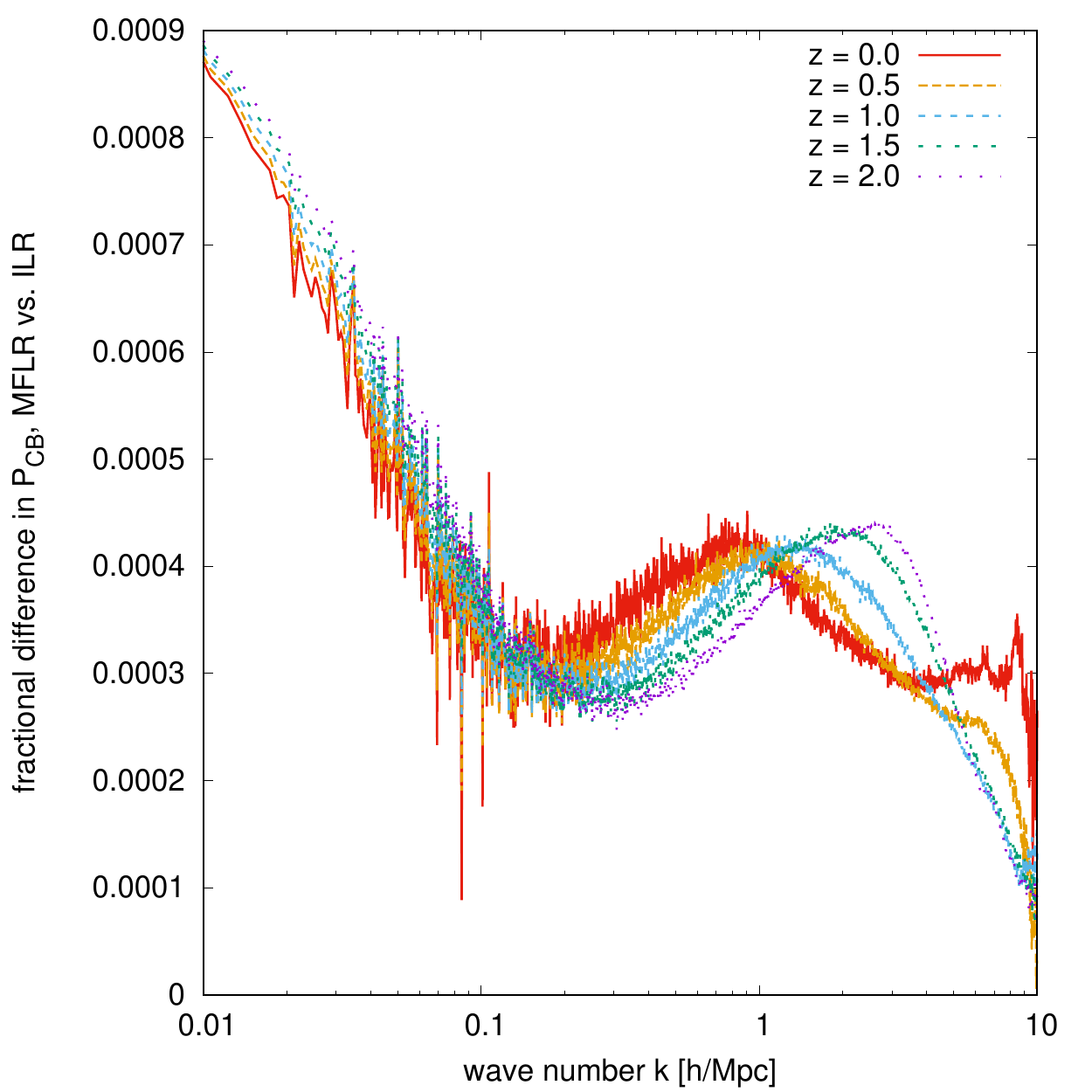}
		\includegraphics[width=75mm]{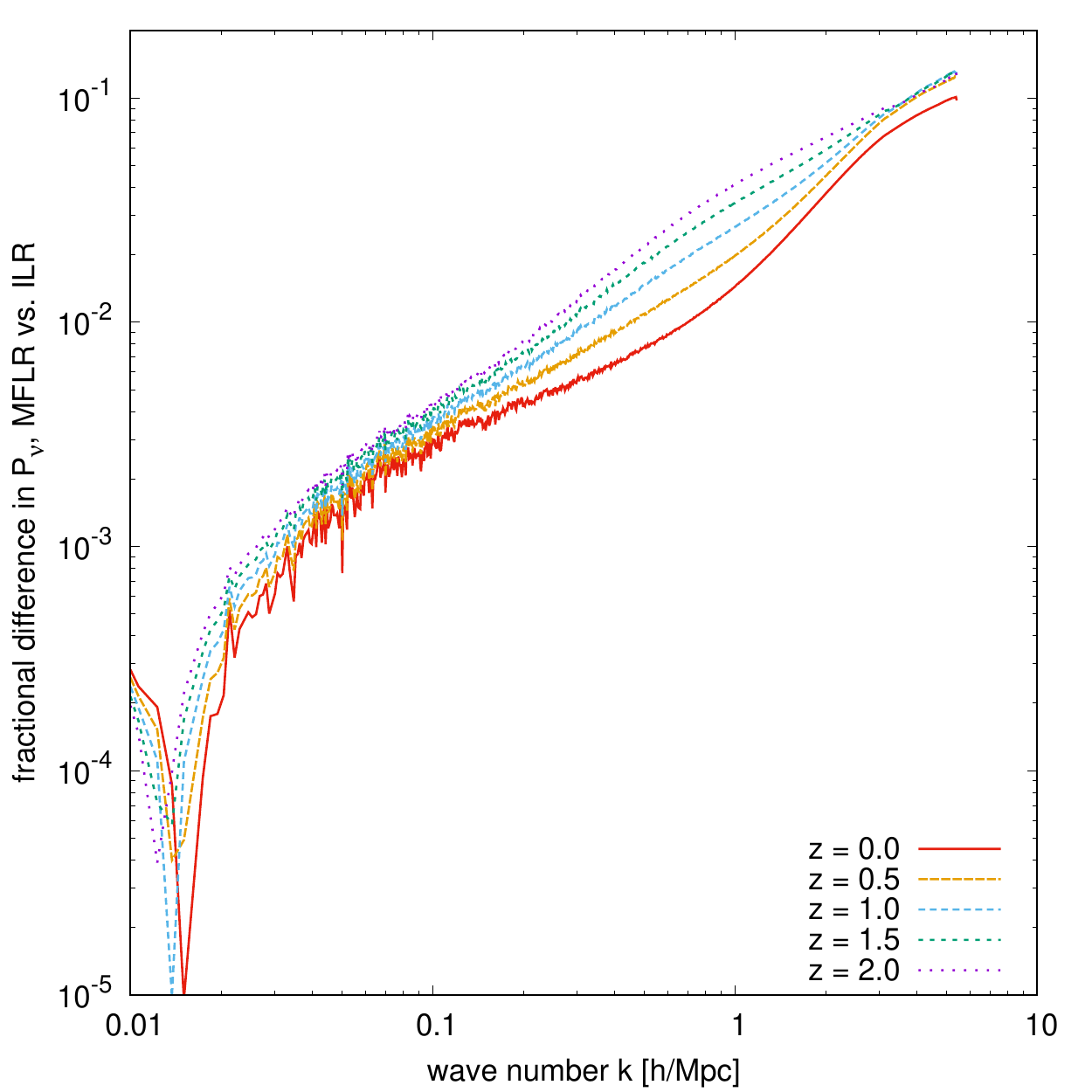}%
	\end{center}
	\caption{Comparison of our  \gadgetcode{} simulations ($L=1024$~Mpc$/h$, $N=1024^3$) using multi-fluid linear response (MFLR)
		and integral linear response (ILR) calculations of  neutrino clustering for the reference $\nu \Lambda$CDM model of table~\ref{t:nuLCDM_reference} at several redshifts.
		{\it Left}:~Ratios of the CDM+baryon power spectra from the two methods.
		{\it Right}:~Ratios of the neutrino power spectra.
		\label{f:gadget2_mflr_ilr}
	}
\end{figure}

Figure~\ref{f:gadget2_mflr_ilr} compares our multi-fluid linear response $N$-body implementation with the integral linear response method of reference~\cite{AliHaimoud:2012vj} (see also section~\ref{subsubsec:ilr}).  From figure~\ref{f:mflr_ilr}, we expect both sets of CDM+baryon and neutrino power spectra to agree at the $0.1\%$ level at the largest scales accessible to our simulation, $k \sim 0.01~h/$Mpc.  At smaller scales, the CDM+baryon power spectra will agree to $\sim 0.01\%$ for sufficiently many neutrino fluids, while the accuracy of the multi-fluid neutrino density will worsen to $\sim 1\%$ at $k > k_\mathrm{FS} \approx 0.2~h/$Mpc.  Figure~\ref{f:gadget2_mflr_ilr} confirms these expectations.  Over a range of redshifts, the CDM+baryon power spectra from both methods agree to $\approx 0.04\%$ at $k \geq 0.05~h/$Mpc, with the difference rising to nearly $0.1\%$ at smaller wave numbers.  On the other hand, the neutrino power spectra shown in the right panel agree to $<1\%$ at $k \leq k_\mathrm{FS}$; on smaller scales, e.g., $k=1~h/$Mpc, however, the agreement worsens to 1.5\% at $z=0$ and $4\%$ at $z=2$.

Thus, we have demonstrated that our modified \gadgetcode{} simulations agree with the literature in terms of their CDM+baryon power spectrum predictions, and that our multi-fluid and integral linear response neutrino calculations return consistent results for both the CDM+baryon and neutrino power spectra.


\subsection{Clustering in the presence of $N$-body CDM+baryon non-linearities}

Having demonstrated convergence and general agreement of our approach with others, we are now in a position to answer quantitatively the three questions we addressed through perturbation theory in section~\ref{subsec:clustering_in_presence_of_cb_NL}:  (i)~How much does linear response to non-linear CDM+baryon clustering enhance neutrino clustering?  (ii)~What fraction of the neutrinos clusters with $\Delta^2 \geq 0.1$, making linear theory unreliable?  (iii)~What effect does this enhancement have on the CDM+baryon power spectrum?  We discuss point~(ii) and then points~(i) and~(iii) in two subsections below.

\subsubsection{Breakdown of neutrino linearity} 
\label{subsec:neutrino_fluids}

\begin{figure}[t]
	\begin{center}
		\includegraphics[width=150mm]{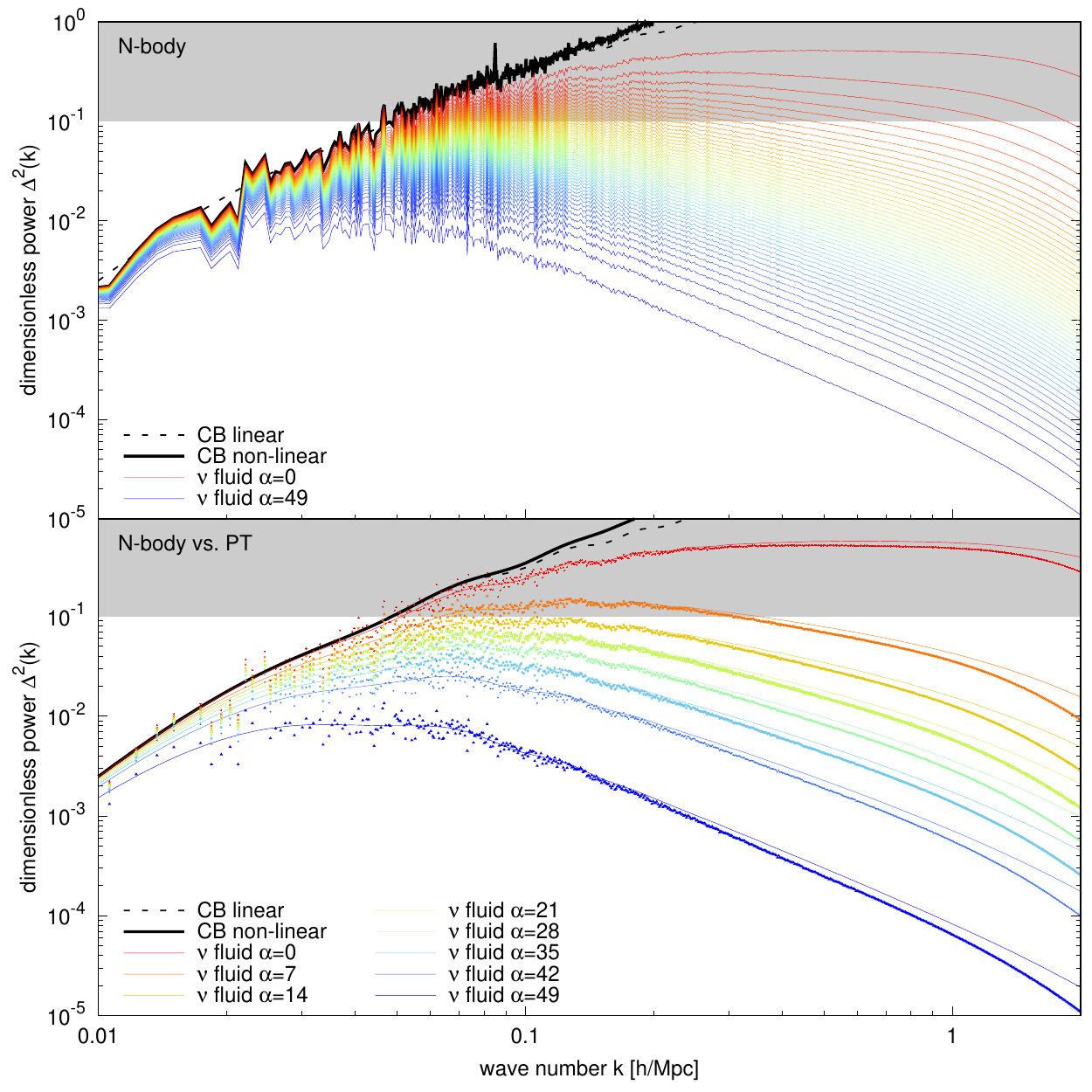}%
	\end{center}
	\caption{Dimensionless $z=0$ CDM+baryon and neutrino monopole power spectra from our multi-fluid linear response simulation of the reference $\nu \Lambda$CDM cosmology, using $L=1024$~Mpc$/h$ and $N=1024^3$, and $N_\tau=50$ and $N_\mu=20$ in the fluid component. {\it Top}:~All fifty neutrino power spectra from the $N$-body simulation,  fourteen of which enter the region $\Delta^2 > 0.1$.
		{\it Bottom}:~Eight representative $N$-body neutrino power spectra (points), together with the corresponding Time-RG perturbative predictions of section~\ref{subsec:clustering_in_presence_of_cb_NL}
		(lines).
		\label{f:nbody_fluids}
	}
\end{figure}

Analogous to in figure~\ref{f:D2_nu}, the top panel of figure~\ref{f:nbody_fluids} shows the $z=0$ neutrino monopole power spectra from our multi-fluid linear response simulation with $N_\tau=50$ neutrino fluids, 
for the reference $\nu\Lambda$CDM model of table~\ref{t:nuLCDM_reference}.
Fourteen of these fluids, or $28\%$ of the neutrino energy density, have dimensionless power exceeding $\Delta^2 = 0.1$.  This result is in close agreement with our Time-RG perturbative calculation from section~\ref{subsubsec:breakdown_of_neutrino_linearity}, where we found $27\%$ of the neutrinos crossing the $\Delta^2 = 0.1$ threshold.  


A direct comparison of the $N$-body and Time-RG perturbative predictions is shown in the bottom panel of figure~\ref{f:nbody_fluids}, where we contrast eight out of the fifty neutrino fluids between the two approaches. Evidently, perturbation theory agrees quite well with simulations in the regime $k \lesssim k_\mathrm{FS} \approx 0.2~h/$Mpc, roughly where the bulk of the neutrino power spectra peak.  
Thus, perturbation theory provides an accurate estimate of the fraction of the neutrino population with $\Delta^2>0.1$, confirming the main result of section~\ref{subsubsec:breakdown_of_neutrino_linearity}.

\subsubsection{Neutrino and CDM+baryon clustering enhancements}
\label{subsec:clustering_enhancement}

\begin{figure}[t]
	\begin{center}
		\includegraphics[width=75mm]{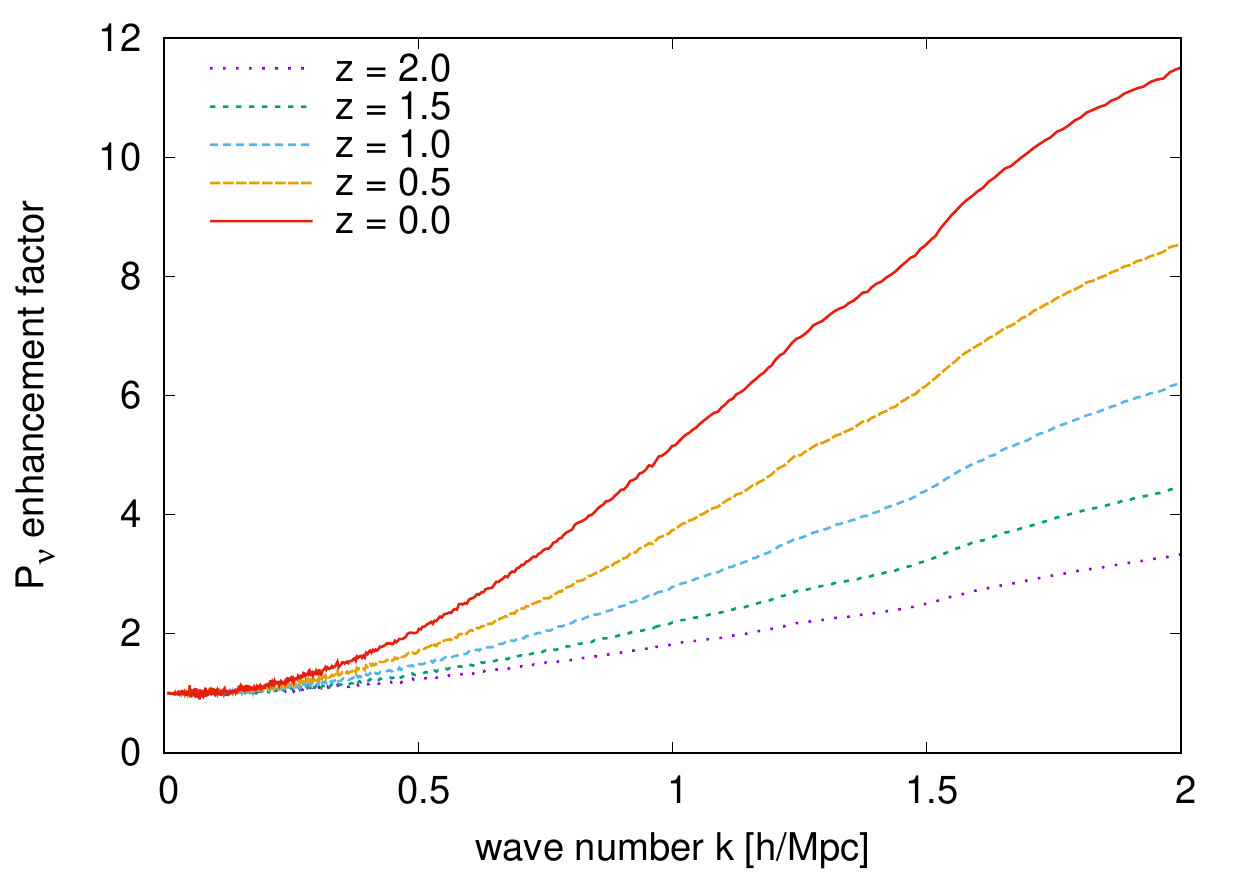}
		\includegraphics[width=75mm]{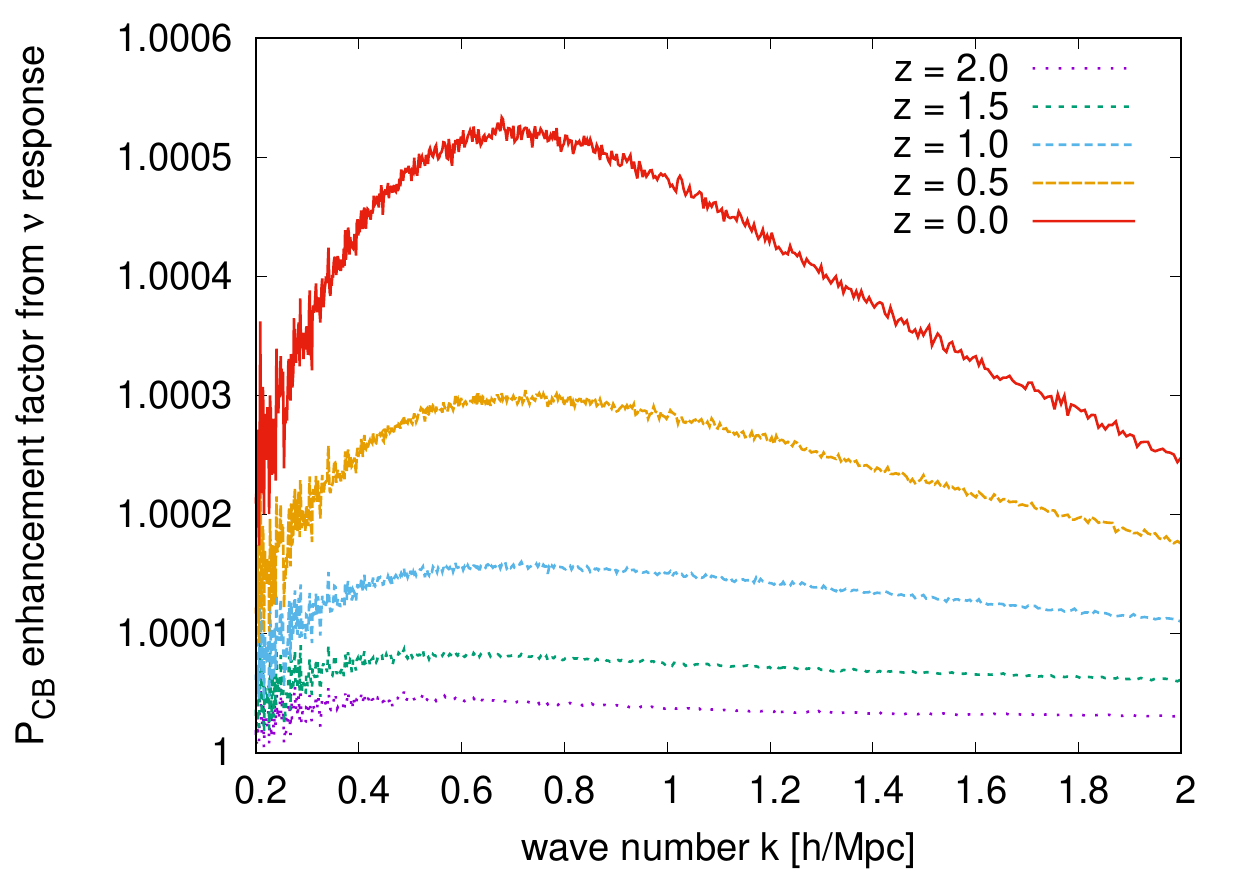}%
	\end{center}
	\caption{Ratios of power spectra computed with multi-fluid $N$-body simulations with and without neutrino linear response to
		non-linear CDM+baryon clustering at various redshifts.
		{\it Left}: Neutrino power spectrum ratios.  Here, enhancement reaches $5.2$ at $z=0$ and $k=1~h/$Mpc, and 
		continues rise up to the resolution limit of our simulation.
		{\it Right}:~CDM+baryon power spectrum ratios.  Enhancement peaks 	at about $0.05\%$ around $k=0.7~h/$Mpc.
		\label{f:nbody_enhancement}
	}
\end{figure}

We conclude this section with figure~\ref{f:nbody_enhancement}, which, analogous to figures~\ref{f:nu_enhancement} and~\ref{f:cb_enhancement}, 
quantifies the enhancement in the  neutrino and CDM+baryon power spectra due to neutrino linear response to non-linear CDM+baryon clustering over a range of redshifts.
Specifically, we compare the outcomes of two simulations, 
one in which neutrinos respond linearly to the non-linear CDM+baryon density perturbations via Poisson's equation,
 and the other in which the ratio $\delta_{\alpha, 0} / \delta_\mathrm{cb}$ for each neutrino fluid $\alpha$ is fixed entirely by linear perturbation theory.

The left panel of figure~\ref{f:nbody_enhancement} shows a neutrino enhancement factor that agrees well with the perturbative predictions shown in figure~\ref{f:nu_enhancement}.  For example, the enhancement factor at $z=0$ and $k=1~h/$Mpc is $5.2$ in our simulation, compared with $5.9$ from perturbation theory.  By $k=2~h/$Mpc, the smallest scale for which our simulations are reliable, neutrino clustering is enhanced by more than an order of magnitude.  At this point, we note that the slowest neutrino fluids that contribute the most to small-scale clustering are also the ones most in need of non-linear corrections.  This means that the actual small-scale neutrino clustering enhancement is very likely larger than our linear response predictions, even in the presence of CDM+baryon non-linearity computed with $N$-body simulations.

In contrast, changes to the CDM+baryon power spectrum due to this enhanced neutrino clustering are modest --- at most $\approx 0.05\%$ at $z=0$ --- as shown in the right panel of figure~\ref{f:nbody_enhancement}.
 This $N$-body estimate is roughly consistent with the perturbative result 
 of section~\ref{subsec:clustering_in_presence_of_cb_NL}, 
  where we found a maximum enhancement of $\approx 0.1$\% at $z=0$.  Considering that our reference cosmology has $\Omno h^2 = 0.01$, i.e., about twice the value of current conservative bounds, we can thus conclude that  CDM+baryon clustering statistics can be computed to $0.05\%$-accuracy or better without taking into account enhanced neutrino clustering.

\section{Conclusions}
\label{sec:conclusions}

We have developed, implemented, and tested a non-relativistic multi-fluid perturbation theory for massive neutrinos.  Beginning with the relativistic, discrete-angle treatment of Dupuy and Bernardeau~\cite{Dupuy:2013jaa,Dupuy:2014vea}, we consider the non-relativistic limit appropriate for late-universe clustering and  describe the angle-dependence of neutrino clustering using a Legendre polynomial expansion with an appropriate truncation.  In figures~\ref{f:Leg_coeff}, \ref{f:mflr_convergence}, \ref{f:mflr_ilr}, and \ref{f:mflr_class}, we demonstrate that the resulting multi-fluid linear perturbation theory converges as the number of fluids and Legendre polynomials is increased, and that it agrees closely with the state-of-the-art neutrino linear response method as well as outputs of the linear  Boltzmann code~\classcode{}.

Using this multi-fluid linear response, we have quantified important aspects of structure formation in massive neutrino cosmologies.  Firstly, the multi-fluid picture is unique in its ability to provide fine-grained information on the clustering of different components of the neutrino velocity distribution.  We take advantage of this rich detail in figures~\ref{f:D2_nu}, \ref{f:vary_Nnu}, \ref{f:vary_wnu}, and \ref{f:nbody_fluids} to quantify the fraction of the neutrino population whose clustering power exceeds $\Delta^2=0.1$, a regime in which linear theory becomes unreliable and non-linear treatments are necessary.  
For a reference $\nu\Lambda$CDM cosmology with $\Omno h^2 = 0.01$, or, equivalently, $\sum m_\nu = 0.93$~eV, split across three equal-mass neutrino species, we find that more than one-fourth of the neutrino population satisfies this non-linearity criterion.  This fraction rises above $90\%$ for a single massive species with $m_\nu = 0.93$~eV, a mass consistent with the sterile neutrino interpretation of the LSND/MiniBooNE/reactor anomalies.
 Even for the smaller $\Omno h^2 = 0.005$ currently still allowed in conservative analyses of cosmological data, $5\%$ of the neutrino population enters this non-linear regime, demonstrating that a non-linear description of neutrino clustering is necessary for percent-level calculations of the neutrino power spectrum.

Furthermore, in figure~\ref{f:nbody_enhancement} we have studied the impact of neutrino linear response on the clustering of CDM and baryons as well as on the neutrinos themselves.  The enhancement of CDM+baryon clustering due to neutrino linear response is modest --- a maximum of $0.05\%$ for neutrino masses currently allowed by cosmological data --- thereby validating approximations  that neglect this enhancement in simulations of the dark matter and galaxy power spectra~\cite{Brandbyge:2008js}. The neutrinos themselves, on the other hand,  are significantly affected by non-linear CDM+baryon clustering.  In our reference $\nu \Lambda$CDM cosmology, the neutrino power spectrum increases by over an order of magnitude at the smallest scales $k \approx 2~h/$Mpc accessible to our simulations.

 In conclusion, 
the multi-fluid linear response approach provides a rich source of information on neutrino clustering that points the way to more accurate non-linear treatments of massive neutrinos in a gravitational setting.  It tells us precisely which neutrino fluids have entered the non-linear regime at which time and provides fine-grained information on the bulk velocity as well as the first-order density and velocity perturbations of each fluid.  With this information, one can envisage a low-noise hybrid simulation scheme in the manner of reference~\cite{Brandbyge:2009ce,Bird:2018all}, but wherein only the slowest neutrino fluids are converted to a particle realisation at as late a time as possible, through which to minimise systematic uncertainties associated with the large neutrino velocity dispersion.  Such innovative approaches are essential as cosmology over the next years seeks to measure, with ever increasing precision and sophistication, the absolute neutrino mass scale, a fundamental parameter of the Standard Model of particle physics, and searches for neutrino physics beyond the Standard Model.


\acknowledgments

JZC acknowledges support from an Australian Government Research Training Program Scholarship.
AU and  Y$^3$W are  supported by the Australian Research Council's Discovery Project (project DP170102382) and Future Fellowship (project FT180100031) funding schemes.   
This research includes computations using the computational cluster Katana supported by Research Technology Services at UNSW Sydney.
We thank Juliana Kwan for useful discussions.


\bibliographystyle{utcaps}
\bibliography{Multi_Fluid}
\end{document}